\documentclass{aa}  
\usepackage[varg]{txfonts}
\usepackage[normalem]{ulem}  
\usepackage{color}
\usepackage{url}
\usepackage{hyperref}

\defcitealias{CAP}{CAP}

\usepackage{upgreek}
\usepackage{graphicx}	
\usepackage{amsmath}	
\usepackage{amssymb}	
\usepackage{xcolor}
\usepackage{comment}

\newcommand{\mdot}{\ensuremath{\dot{m}}}
\newcommand{\ergl}{\ensuremath{\rm erg\, s^{-1}}}
\newcommand{\alert}[1]{{\color{red}#1}}

\DeclareRobustCommand{\uppartial}{\text{\rotatebox[origin=t]{20}{\scalebox{0.95}[1]{$\partial$}}}\hspace{-1pt}}

\begin{document} 
\title{Super-Eddington accretion discs with advection and outflows around magnetized neutron stars}
\titlerunning{Super-Eddington accretion discs  around neutron stars}

 \author{Anna Chashkina
          \inst{1,2} 
          \and
          Galina Lipunova\inst{2}
          \and
          Pavel Abolmasov\inst{1,2}
          \and
          Juri Poutanen\inst{1,3,4}}

   \institute{Tuorla Observatory, Department of Physics and Astronomy,  FI-20014 University of Turku, Finland \\
   \email{anna.chashkina@utu.fi, juri.poutanen@utu.fi}
   \and
   Sternberg Astronomical Institute, Lomonosov Moscow State University, Universitetsky pr. 13, 119992 Moscow, Russia
   \and
   Space Research Institute of the Russian Academy of Sciences, Profsoyuznaya str. 84/32, 117997 Moscow, Russia
   \and Nordita, KTH Royal Institute of Technology and Stockholm University, Roslagstullsbacken 23, SE-10691 Stockholm, Sweden
}

\abstract{We present a model for a super-Eddington accretion disc around a magnetized neutron star taking into account advection of heat and the mass loss by the wind. 
The model is semi-analytical and predicts radial profiles of all  basic physical characteristics of the accretion disc. 
The magnetospheric radius is found as an eigenvalue of the problem. 
When the inner disc is in radiation-pressure-dominated regime but does not reach its local Eddington limit, advection is mild, and the radius of the magnetosphere depends weakly on the accretion rate.  
Once approaching the local Eddington limit, the disc becomes advection-dominated, and the scaling for the magnetospheric radius with the mass accretion rate is similar to the classical Alfv\'en relation. Allowing for the mass loss in a wind leads to an increase of the magnetospheric radius. 
Our model may be applied to a large variety of magnetized neutron stars accreting close to or above their Eddington limits: ultra-luminous X-ray pulsars, Be/X-ray binaries in outbursts, and other systems. In the context of our model we discuss the observational properties of NGC\,5907~X-1, the brightest ultra-luminous pulsar known so far, and  NGC\,300~ULX-1 which is apparently a Be/X-ray binary experiencing a very bright super-Eddington outburst. 
}


   \maketitle


\section{Introduction}\label{sec:introduction}

Mass transfer rates in binary systems may vary in very broad limits, from very low to the amounts vastly exceeding the Eddington limit of the accretor. Highly super-Eddington accretion rates are not surprising in binary systems containing a neutron star (NS) and a massive star filling its Roche lobe.
In this case, the mass transfer runs on a relatively short time scale, being the thermal time scale of a tens-solar-mass star, which is millions of years.
The mass transfer does not quench because the large mass ratio makes a binary system more likely to remain in contact  or even to tighten, thus increasing the mass transfer rate \citep[see e.g.][]{Ivanova15,Ivanova17}.

The strong magnetic field makes it possible to transport the accreted matter deep into the gravitational well to the NS surface. 
{ Presumably,} the matter falls to the NS surface in a thin curtain (see fig. 1a in \citealt{BS76}). Large ratio of the radiating surface area to the volume as well as the reduced scattering cross-section in the strong magnetic field  { are invoked to explain the observed excess above the Eddington limit \citep[see, e.g.,][]{2015MNRAS.454.2539M,Kawashima+2016}.}

Recently it was realised that some of ultraluminous X-ray sources (ULX) in the nearby galaxies are actually accreting magnetized NSs. 
\citet{bachetti14} using \textit{NuSTAR} data discovered coherent pulsations in the ULX X-2 in the galaxy M82.  
Later, two more similar objects, NGC\,7793~P13 \citep{Israel_7793,2016ApJ...831L..14F} and NGC\,5907~X-1\ \citep{Israel_5907} were found. 
NGC\,5907~X-1\ is exceptional in its luminosity, exceeding $10^{41}\ergl$ during some of the observations.
These are the prototypical members of the ULX pulsar (ULXP) family. 
More recently, other ULXPs have been found.
Pulsations with 20--30\,s period were discovered in NGC\,300 ULX1, the supernova impostor SN2010da \citep{villar16}, with a very strong spin up of $\dot{p}=-1.75 \times 10^{-7}$\,s\,s$^{-1}$ \citep{Bachetti2018}. 
Also, M51 ULX-7 was identified with a ULXP (G. Israel, priv. communication), and, finally, the first ULXP in the Milky Way, Swift~J0243.6+6124, was discovered  \citep{2017ATel10809....1K,2018MNRAS.479L.134T,2018ApJ...863....9W}. 
Though only few persistent super-Eddington objects are robustly identified so far as NSs, it is quite natural  to expect more supercritically accreting NSs among the more general class of ULXs \citep[see][]{Kaaret17}. 

Studies of super-Eddington accretion discs  started simultaneously with the standard disc theory by \citet{SS73}. 
They suggested that, for the accretion rate $\dot M_{0}$ exceeding a certain critical value $\dot M_{\rm cr}$, a wind emanates inside the spherization  radius $R_\mathrm{sph}$, driven by the radiation pressure, and removes just the right amount of matter to keep the disc at the  {  local} Eddington limit. 
The luminosity of the supercritical disc in this case exceeds the Eddington luminosity $L_{\rm Edd}$ by a logarithmic factor $1+ \ln (\dot M_{0}/\dot M_{\rm cr})$,
and the accretion rate decreases with radius as $\dot M(R)=\dot M_{0}\,R/R_{\rm sph}$. 
We will refer to such a scenario as `classical mass loss'. Implications of this scenario to the discs around magnetized NSs were considered by \citet{lipunov82} and recently by  \citet{2017AstL...43..464G}.
Remarkably, because the properties of the inner disc in this model are independent of the outer boundary conditions, the size of the magnetosphere and the luminosity at high $\dot M_{0}$ converge to universal values dependent on the NS magnetic moment $\mu$ only. 

Later it was realized that it is important to take into account other effects related to deviations from the thin disc approximation. 
The most significant departure from the standard disc model is the heat advection, that violates the locality assumption in the  energy balance of the disc. {  The role of advection for the case of optically thin discs was considered, for example, by \citet{ichimaru77},  \citet{narayan94,narayan95}, \citet{Abramowicz95,Abramowicz96}, and for the optically thick case by \citet{1982ApJ...253..873B} and \citet{Slim88}. 
Numerical simulations of supercritical accretion onto a black hole were performed by \citet{ECK88}, {\citet{Beloborodov1998}}, \citet{OM05}, \citet{OTT05}, and more recently by \citet{2011ApJ...736....2O}, \citet{mckinney14}, \citet{sadowski14}, \citet{sadowski16}, and \citet{2017PASJ...69...33O}.  
Semi-analytical models for supercritical discs including heat advection and outflows were constructed by  \citet{lipunova}, \citet{Kit02}, \citet{Fuk04}, and \citet{poutanen07}. 
}

{  While most of this theoretical work on supercritical accretion was devoted to discs around black holes, the discovery of ULXPs has drawn attention to magnetized NSs accreting at high rates. Recently, simulations of supercritical accretion onto a non-magnetized NS were performed by \citet{abarca2018}.}
It is important to note that even in highly supercritical ULXPs like M82~X-2 and NGC 7793 P13, the accretion disc outside the magnetosphere may remain in a sub-critical regime.
\citet{CAP} (hereafter CAP) considered accretion onto a magnetized NS in a regime that may be characterized as intermediate, with the accretion rate in the range $(10^{-9}-10^{-6}){\rm \,M_\odot\,yr^{-1}}$ for $\mu \sim 10^{31}- 10^{32} {\rm \,G\, cm^3}$ inferred for ULXPs (see \citealt{Tsygankov16}). The disc outside the magnetosphere in this case remains geometrically thin, nearly-Keplerian, and does not lose considerable amounts of matter.
This allowed us to use certain results of the standard disc theory such as local radiation energy balance and simple conservation laws for mass and angular momentum. 
This approach works well if the formally calculated spherization radius is smaller than the Alfv\'{e}n radius. However, super-Eddington NSs  with  lower magnetic moments or accreting at larger rates should possess a supercritical advective disc, whose radiation pressure becomes sufficient to unbind part of the accreted material.

In the present work, we develop a model of a supercritical accretion disc around a magnetized NS, applicable to a broader class of objects, including NSs with pulsar-scale magnetic fields and luminosities of hundreds of the Eddington limit, and, in particular,  extremely bright sources like NGC\,5907~X-1. 
The structure of the paper is as follows.  
In Sect.~\ref{sec:model} we describe our model and present the main equations describing disc accretion accounting for the effects of advection and the mass and momentum loss in a wind. 
Sect.~\ref{sec:res} is devoted to the results, where we present the dependencies of the magnetospheric radius on the mass accretion rate and NS magnetic moment. We also consider  the effects of the pulsar spin and the irradiation of the disc by the central source on the disc structure. 
We discuss our results and apply them to particular ULXPs in Sect.~\ref{sec:discussion}.
We conclude in Sect.~\ref{sec:conclusion}.

\begin{figure}
\includegraphics[width=\columnwidth]{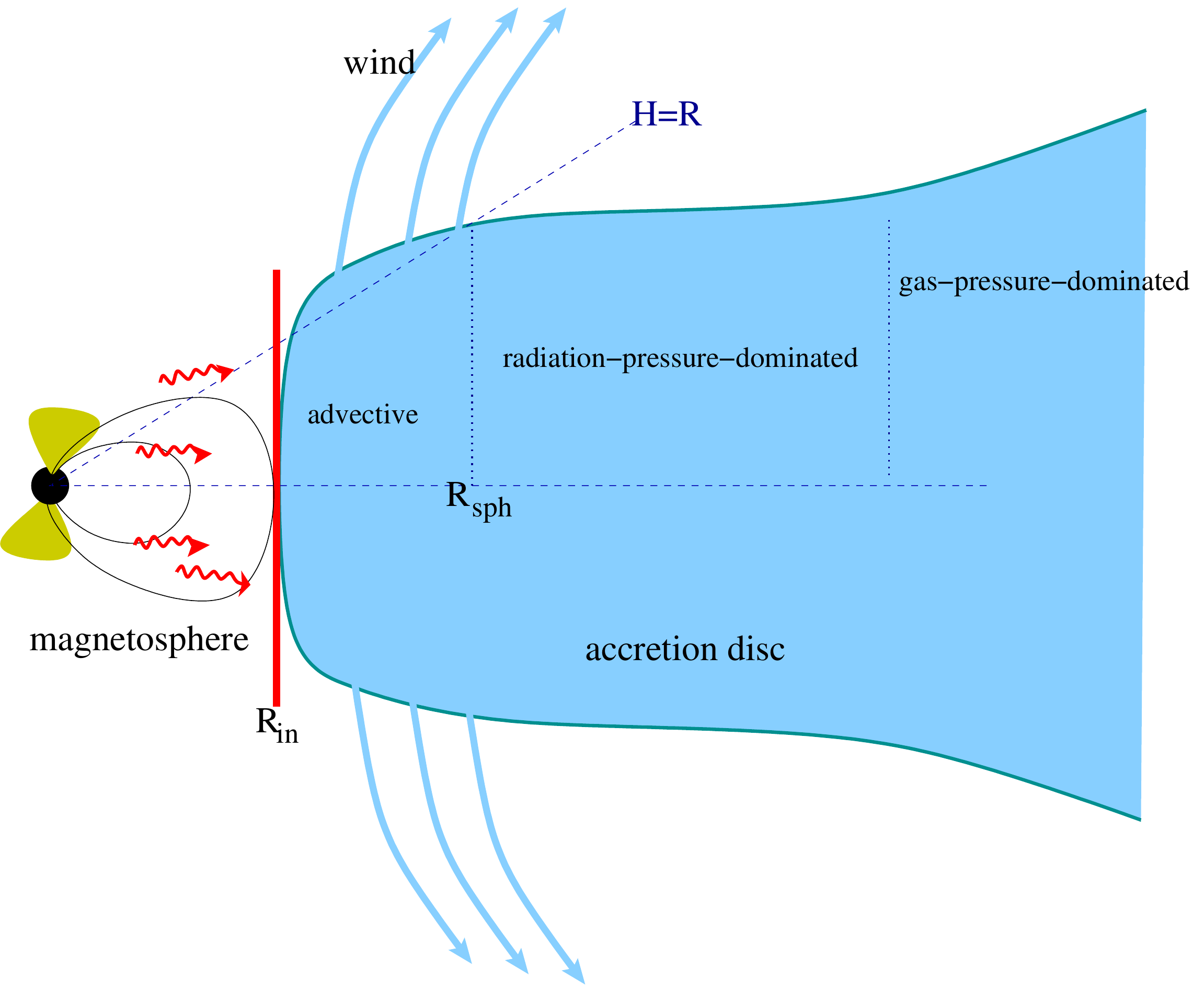}
\caption{Structure of an accretion disc around a ULXP. For very high mass accretion rates, the inner parts of the disc, inside the  spherization radius $R_{\rm sph}$ indicated in the sketch, enter the super-Eddington accretion regime. Inside $R_{\rm sph}$, the  thin disc model is not applicable. Mass loss in a wind is shown by blue arrows. The accretion column, where most of the energy is released, is shown by yellow cones, and the red wavy lines refer to the radiation of the column that may affect the inner disc pressure balance. The red vertical line marks the effective boundary between the disc and the magnetosphere at $R_{\rm in}$.} \label{fig:disc}
\end{figure}

\section{Model}\label{sec:model}
\subsection{Basic equations}

We will consider a NS of mass $M$, radius $R_{\rm NS}$ and magnetic moment $\mu$. We assume it is an aligned rotator with the unperturbed dipolar magnetic field inside the magnetospheric radius, which is equal to the inner radius of the disc $R_{\rm in}$. Outside this radius, the magnetic field lines are opened by the ideally conducting accretion disc in the way it was proposed by  \citet{parfrey}. All the interactions between the disc and the magnetosphere are assumed to occur in a narrow strip at the edge of the disc and are described by only two boundary conditions introduced in \citetalias{CAP} and later in this section.  {  A simple sketch of the adopted structure of the disc is shown in Fig. \ref{fig:disc}.}

In our model, we include three effects important for an accretion disc at near- or super-critical accretion rate: (i) advection, (ii) mass  loss in the wind, and (iii) angular momentum losses in the wind. Advection is related to the increasing photon diffusion time scales at large mass accretion rates. It alters the energy equation. The matter loss in the wind makes the disc accretion rate dependent on radius and also affects the angular momentum and the energy conservation equations because the wind carries some angular momentum and energy. \citet{poutanen07} assumed that the specific angular momentum of the wind is equal to that of the matter in the disc. 
However, if the wind is magnetized, it can remove efficiently the angular momentum from the disc, as in a centrifugally-driven wind model of \citet{BP82}.

We follow the basic framework of \citetalias{CAP} and retain some of the important assumptions of the model. 
{  First of all, the boundary conditions at the inner disc edge remain the same.
We take the torque balance in the form}
\begin{equation}\label{eq:base:wbalance}
\dot{M}_{\rm in}\, (\Omega_{\rm in}-\Omega_{\rm NS})\,R^2_{\rm in}=k_{\rm t}\,\displaystyle\frac{\mu^2 H_{\rm in}}{R^4_{\rm in}}+L\,\displaystyle\frac{\Omega_{\rm in}}{c^2}H_{\rm in}R_{\rm in},
\end{equation}
and {  assume the pressure balance, which yields for the viscous stress tensor
the following relation}
\begin{equation}\label{eq:base:pbalance}
W^{\rm in}_{r\phi}=2\alpha H_{\rm in} \left( \displaystyle\frac{\mu^2}{8\uppi R_{\rm in}^6}+\displaystyle\frac{L}{4\uppi R_{\rm in}^2 c}\right)  .
\end{equation}
Here, $\dot M_{\rm in}$ is the mass accretion rate at the inner edge of the disc, $\Omega_{\rm NS}$ is the angular velocity of the NS, $\Omega_{\rm in}$, $H_{\rm in}$, and $W_{r\phi}^{\rm in}$ are the angular velocity, the half-thickness and the vertically integrated $r\phi$- component of the viscous stress tensor at the inner boundary of the disc $R_{\rm in}$,  the dimensionless constant $k_{\rm t}$ 
{  parametrizes the efficiency of the angular momentum removal by the magnetic and viscous torques, and}
{  $\alpha$ is dimensionless viscosity parameter as defined by \citet{SS73}}. The luminosity $L=\eta\, \dot{M}_{\rm in}\, c^2$ is  released close to the NS in the accretion column with  efficiency $\eta$ as measured by an imaginary observer in the equatorial plane at the inner edge of the disc. Because the accretion column emits anisotropically, the efficiency $\eta$ may differ from the angle-averaged efficiency of an isotropic source as well as from the efficiency measured by an observer at infinity. The radial structure of the disc is described by the angular frequency $\Omega$, thickness $H$, and the viscous stress $W_{r\phi}$. The angular frequency and the viscous stress should conform to the boundary conditions~(\ref{eq:base:wbalance}) and (\ref{eq:base:pbalance}). 
{   Equation (\ref{eq:base:wbalance}) is similar to the inner boundary condition in \citet{Spruit93} and \citet{Spruit2004} in the case of a slowly rotating NS. 
The additional term in the RHS is radiation drag that can be important at large luminosities $L>L_{\rm Edd}$. Note that equation (\ref{eq:base:wbalance}) is valid only for the case of the disc rotating faster than the magnetosphere. For the propeller case ($\Omega_{\rm in}<\Omega_{\rm NS}$), the boundary condition would be different. Moreover, one could consider a radial distribution of the magnetic torque applied to the disc, as done, for example, by \citet{KR07}. On the other hand, \cite{Matt-Pudritz2005} argue that if $\Omega_{\rm in}>\Omega_{\rm NS}$, the magnetic interaction can be limited to a narrow ring. 
}

We take the radial component of the momentum equation in the form
\begin{equation}\label{eq:base:omega}
\Omega^2 R=\displaystyle\frac{1}{\Sigma}\frac{\uppartial \Pi}{\uppartial R}+\displaystyle\frac{GM}{R^2},
\end{equation}
where $\Sigma=\displaystyle\int^H_{-H} \rho \, {\rm d}z$ is the surface density and $\Pi=\displaystyle\int^H_{-H} P \,{\rm d}z$ is the vertically-integrated pressure. 
As in \citetalias{CAP}, we ignore the dynamic term $v_R \,{\rm d} v_R/{\rm d} R$ which is suppressed by a factor of $\alpha^2 (H/R)^2\ll 1$ with respect to the pressure gradient term. 

In all the equations for the radial structure, we use vertically integrated quantities. In the equations for the vertical structure we assume a fixed vertical effective polytropic index $n$, defined through the equation for the density $\rho = \rho_{\rm c}[ 1-(z/H)^2]^n$ and for the pressure $P=P_{\rm c} [ 1-(z/H)^2]^{n+1}$ \citep{Paczynski78}.  
This allows us to compute analytically the connection factors between the midplane quantities  such as $\rho_{\rm c}$, $P_{\rm c}$, and the corresponding vertically-integrated values $\Sigma$, $\Pi$ as  described in Appendix~\ref{sec:vert}.
For the vertical temperature profile used in the advection term calculations, we use the expression $T= T_{\rm c}[ 1-(z/H)^2]^{(n+1)/4}$ valid for the radiation-pressure dominated case, as $T\propto p^{1/4}$, here $T_{\rm c}$ is the central temperature on the disc. 

The angular momentum conservation equation is modified by an additional term corresponding to the angular momentum outflow in the wind:
\begin{equation}\label{eq:base:angmo}
\displaystyle\frac{{\rm d}\left(\dot M(R) \Omega R^2\right)}{{\rm d}R}=\displaystyle\frac{{\rm d}}{{\rm d}R}\left( 2\uppi R^2 W_{r\phi}\right) +\displaystyle\frac{{\rm d}\dot M(R)}{{\rm d}R}\Omega R^2 \psi,
\end{equation}
where $\psi\geq 1$ allows us to scale up the net angular momentum lost in the wind. If $\psi=1$, the net angular momentum in the wind is equal to that in the disc \citep{lipunova,poutanen07}. Larger $\psi$ can appear in magneto-centrifugal winds, where the sub-Alfvenic part of the transonic flow rotates approximately rigidly, thus increasing the angular momentum loss by a factor of $\sim (R_{\rm wA}/R)^2$, where $R_{\rm wA}$ is the cylindrical radius of the Alfv\'{e}n surface, and $R$ is the launch point of the wind streamline (see also \citealt{MP05} who used this approach for magnetospheric outflows). 
In the centrifugally driven wind model of \citet{BP82}, for instance, $R_{\rm wA}/R$ may be as large as 5. Simulations of the centrifugally driven winds  \citep[e.g.][]{ustyugova}, {Lovelace1995} suggest even larger lever lengths at low accretion rate, while at high accretion rates, when the wind is  radiatively driven, the effect of magnetic stresses is smaller \citep[see e.g.][]{proga00}. 

The mass accretion rate derivative ${\rm d}\dot{M}/{\rm d}R$ needs to be calculated using some additional equations describing the physics of the wind launching. 
We assume that some fraction $\epsilon_{\rm w}\leq 1$ of the energy leaving the disc with radiation is spent to accelerate the outflow \citep{lipunova,poutanen07}:
\begin{equation}\label{eq:base:ewind}
\epsilon_{\rm w} Q_{\rm rad}=\epsilon_{\rm w} 2\sigma_{\rm SB}T^4_{\rm eff} =\displaystyle\frac{\Omega^2_{\rm K} R}{4\pi}\displaystyle\frac{{\rm d}\dot M(R)}{{\rm d}R},
\end{equation}
where $\Omega_{\rm K}$ is Keplerian angular velocity and $T_{\rm eff}$ is effective temperature. 
The model of optically-thick, energy driven wind developed by \citet{poutanen07} gives similar results. 
Physically, there could be wind losses everywhere in the disc, but we take into account only the continual radiation driven wind which would operate within the spherization radius $R_{\rm sph}$ defined by the condition on the minimal relative thickness $H/R>\left(H/R\right)_{\rm cr}$. 
Because the details of wind launching are uncertain, we assume $\left(H/R\right)_{\rm cr}=1$. 
Note that for a disc around a black hole, the wind mass loss is determined only by the mass of the black hole and the outer accretion rate. For a  disc around a magnetized neutron star, the mass lost in the wind depends also on the magnetic moment of the star and its spin period. 

We use the $\alpha$-viscosity prescription for the vertically integrated viscous stress 
\begin{equation}\label{eq:base:alpha}
W_{r\phi}=\alpha \Pi.
\end{equation}
The midplane pressure may be expressed as a sum of the radiation and gas pressures:
\begin{equation}\label{eq:base:pressure}
\displaystyle\frac{1}{G_{n+1}}\displaystyle\frac{W_{r\phi}}{\alpha H}=\displaystyle\frac{aT_{\rm c}^4}{3}+\frac{1}{G_n}\displaystyle\frac{\Sigma k T_{\rm c}}{H {\tilde{m}}},
\end{equation}
where $\tilde{m}$ is the mean particle mass and  $G_n=\displaystyle\int^1_{-1}(1-x^2)^n {\rm d}x$, see equation~(\ref{eq:vert:factorn}).
Here we have taken into account the $\alpha$-viscosity prescription and the relations between the central and vertically-integrated quantities implied by the adopted vertical structure (equations \ref{eq:vert:Sigma} and \ref{eq:vert:Pi}). We will not use equation~(\ref{eq:base:pressure}) directly, but as a supplement to the energy equation (see Sects.~\ref{sec:adve} {  and \ref{sec:solv})}. Treatment of advection is a  new part of the model that requires a separate consideration. 

\subsection{Advection}\label{sec:adve}
  
In general case with advection, some fraction of energy is radially transported. The energy flux carried by radiation diffusion in the vertical direction is no more equal to the local energy release: 
\begin{equation}\label{eq:energy_total}
Q^{+}=Q_{\rm rad}+Q_{\rm adv},
\end{equation}
where the total energy released in the disc at radius $R$ per unit area is 
\begin{equation}\label{eq:qplus}
Q^+ = W_{r\phi} R \left|\frac{{\rm d}\Omega}{{\rm d}R}\right|,
\end{equation}
and the radiation flux from both sides of the disc is (see equation~\ref{eq:vert:qrad}):
\begin{equation}\label{eq:rad_term}
Q_{\rm rad}=2\sigma_{\rm SB}T^4_{\rm eff}=\displaystyle\frac{16}{3\kappa \Sigma}\,(n+1)\,G_n \sigma_{\rm SB} T_{\rm c}^4 .
\end{equation}
Here $\kappa\simeq 0.34\, $cm$^2$g$^{-1}$ is the Thomson scattering opacity.
The advected flux $Q_{\rm adv}$ may be viewed (as it is done, for instance, in~\citealt{lipunova}) as the flux of heat carried with the flow, and thus may be expressed through the specific entropy per particle~$s$ 
\begin{equation}\label{eq:adv:gene}
 Q_{\rm adv}=\int_{-H}^{H} \rho v_{R}\,\frac{kT}{{\tilde m}} \displaystyle\frac{{\rm d}s}{{\rm d}R}\,{\rm d}z\, ,
\end{equation}
where the radial velocity (taken by its absolute value) is $v_R=\dot M/(2\pi R \Sigma)$. 
{  Taking into account equation (\ref{eq:base:pressure}) we can expand the expression above as (see Appendix~\ref{sec:appendix2}) }
\begin{eqnarray}\label{eq:qadv}
\displaystyle Q_{\rm adv}&=&-\frac{1}{2\pi (n+1)} \frac{1}{G_n} \frac{\dot M W_{r\phi}}{R\Sigma \alpha}
\left[\frac{{\rm d}\ln \Sigma}{{\rm d}R}{\cal S}+\frac{{\rm d}\ln W_{\rm r \phi}}{{\rm d}R}{\cal P}\right. \nonumber\\
\displaystyle  &+&\left. \frac{{\rm d}\ln T_{\rm c}}{{\rm d}R}{\cal Q}+\frac{3}{2R}{\cal R}\right],
\end{eqnarray}
where the coefficients ${\cal S}$, ${\cal P}$, ${\cal Q}$, and ${\cal R}$ are given by equations~(\ref{eq:derive:S})--(\ref{eq:derive:R}). 
The energy balance can be written using equations (\ref{eq:energy_total}), (\ref{eq:qplus}) and (\ref{eq:qadv}) as:
\begin{eqnarray}\label{eq:advection}
\lefteqn{\displaystyle W_{r\phi } R\displaystyle\left| \frac{{\rm d}\Omega}{{\rm d}R}\right| 
 =  \displaystyle\frac{16}{3\kappa \Sigma}\,(n+1)\,G_n \sigma_{\rm SB} T_{\rm c}^4 
-\displaystyle\frac{1}{2\pi(n+1)} \frac{1}{G_n}\frac{\dot M W_{r\phi}}{\alpha R\Sigma} } 
\nonumber \\
&\times& \left[\frac{{\rm d}\ln \Sigma}{{\rm d}R}{\cal S}+\frac{{\rm d}\ln W_{\rm r \phi}}{{\rm d}R}{\cal P}  +\frac{{\rm d}\ln T_{\rm c}}{{\rm d}R}{\cal Q}+\frac{3}{2R}{\cal R}\right] .
\end{eqnarray}
This relation may be re-written for the derivative of the angular velocity as
\begin{equation}\label{E:oomega}
\displaystyle\left|\frac{{\rm d}\Omega}{{\rm d}R}\right|=C_{\Omega}
-\frac{{\rm d}\ln \Sigma}{{\rm d}R}C_{\Sigma}
-\frac{{\rm d}\ln W_{\rm r\phi}}{{\rm d}R}C_{\rm wrf}
-\frac{{\rm d}\ln T_{\rm c}}{{\rm d}R}C_{\rm T}-C_{\rm free}\, ,
\end{equation}
where the coefficients $C_{\Omega, \Sigma, w,  {\rm T, free}}$ are  given by expressions~(\ref{eq:comega})--(\ref{eq:cfree}).

\subsection{Solving the disc equations}\label{sec:solv}

Unlike the previous work \citepalias{CAP}, the number of variables  we need to solve differential equations for is five: the angular frequency $\Omega$, the tangential stress $W_{r\phi}$, the surface density $\Sigma$, the mass accretion rate $\dot{M}$, and the midplane temperature $T_{\rm c}$. 
There are also several adjustable free parameters: viscosity parameter $\alpha$, wind efficiency parameter $\epsilon_{\rm w}$, wind magnetization parameter $\psi$, polytropic index $n$, spin period $p$ and the  magnetic moment of a NS $\mu$, and the accretion efficiency $\eta$. 
We calculate the disc structure from its outer edge to the inner boundary. 
 We solve the five equations listed below:
\begin{enumerate}
\item The derivative of vertically-integrated viscous stress that can be obtained from the radial Euler equation  (\ref{eq:base:omega}) and $\alpha$-viscosity prescription~(\ref{eq:base:alpha}): 
\begin{equation}\label{eq:dif:dwrf}
\displaystyle\frac{{\rm d}W_{r\phi}}{{\rm d}R}=\alpha\Sigma\left(\Omega^2R-\frac{GM}{R^2} \right).
\end{equation}
\item Whenever the outflow condition ($H/R>(H/R)_{\rm cr}$) is satisfied, the mass accretion rate changes with radius according to equation  (\ref{eq:base:ewind})
\begin{equation}\label{eq:dif:massacre}
\displaystyle\frac{{\rm d} \dot M}{{\rm d}R}=\epsilon_{\rm w} \frac{8\,\pi\, \sigma_{\rm SB}\,T^4_{\rm eff}}{\Omega^2_{\rm K} R}=\epsilon_{\rm w}\, \displaystyle\frac{64\,\pi\, \sigma_{\rm SB}}{3\,\kappa\, \Sigma\,\Omega^2_{\rm K}R}\,(n+1) \,G_n\,T_{\rm c}^4.
\end{equation}
\item The differential equation for the central temperature may be obtained by taking derivative of the pressure equation~(\ref{eq:base:pressure})
\begin{eqnarray}\label{eq:dif:tcent}
\displaystyle\frac{{\rm d}\ln T_{\rm c}}{{\rm d}R} 
&=& (8-6\beta)^{-1}\left[\frac{\Sigma \alpha (1+\beta)}{W_{r\phi}} \left(\Omega^2 R-\displaystyle\frac{GM}{R^2} \right)
 \right. \nonumber \\
&+& \left. (1-3\beta)\frac{{\rm d}\ln \Sigma}{{\rm d}R} 
- \displaystyle\frac{3(1-\beta)}{R}\right].
\end{eqnarray}
Here, we used the thickness of the disc $H$ following from the hydrostatic equilibrium (see equation~\ref{eq:vert:h}) together with the gradient of $W_{r\phi}$ substituted from equation~(\ref{E:oomega}). We also use the ratio of the gas pressure to the total pressure in the equatorial plane expressed using the adopted vertical structure (see equation~\ref{eq:vert:beta})
\begin{equation}\label{eq:base:beta}
\beta=\frac {P_{\rm g}}{P_{\rm tot}} = \frac{2(n+1)}{2n+3}\frac{\alpha \Sigma kT_{\rm c}}{\tilde{m} \Pi}\, . 
\end{equation}

\item To determine the radial dependence of the angular velocity, we substitute equations~(\ref{eq:dif:dwrf}) and (\ref{eq:dif:massacre}) into angular momentum conservation equation (\ref{eq:base:angmo}):
\begin{eqnarray}\label{eq:dif:omega}
\displaystyle\frac{{\rm d}\Omega}{{\rm d}R}
&=&\frac{2\pi \Sigma \alpha}{\dot M}\left(\Omega^2 R-\frac{GM}{R^2}\right)  \\
&+&\epsilon_{\rm w}(\psi-1) \frac{64\pi}{3}(n+1)G_n\frac{ \sigma_{\rm SB} T^4_{\rm c}\Omega }{\kappa\Sigma \dot M R \Omega^2_{\rm K}}
 -\displaystyle\frac{2\Omega}{R}+\frac{4\pi W_{r \phi}}{\dot M R}. \nonumber 
\end{eqnarray}
The second term on the right-hand side is switched off for a thin disc ($H/R<(H/R)_{\rm cr}$) when there no outflows. 

\item The differential equation for the surface density was obtained from the advection equation (\ref{E:oomega}) substituting all other derivatives from equations (\ref{eq:dif:dwrf})--(\ref{eq:dif:omega}):
\begin{eqnarray}\label{eq:dif:sigma}
\displaystyle\frac{{\rm d}\ln \Sigma}{{\rm d}R}&=& \left[C_{\rm \Sigma}+C_{\rm T}\frac{1-3\beta}{8-6\beta}\right]^{-1}
\displaystyle \left[ 
C_{\rm \Omega}-\displaystyle\left|\frac{{\rm d}\Omega}{{\rm d}R}\right|\right. \nonumber\\
&-&\displaystyle\frac{\alpha \Sigma}{W_{r\phi}} \left(\Omega^2 R- \frac{GM}{R^2}\right) \left(C_{\rm wrf}+C_{\rm T}\frac{1+\beta}{8-6\beta}\right) \nonumber \\
&+&\left. \displaystyle\frac{3C_{\rm T}}{R}\frac{1-\beta}{8-6\beta}-C_{\rm free} 
\right].
\end{eqnarray}
\end{enumerate}
The disc structure equations are further converted to a more compact dimensionless form in Appendix~\ref{sec:notation}.

For two of the variables, $\Omega$ and $W_{r\phi}$, the boundary conditions exist that need to be satisfied. All the quantities at the outer boundary of the integration region should conform to the thin disc model with $\dot{M}=\dot{M}_{0}$ being one of the global parameters of the simulation and $W_{r\phi}=W_{r\phi}^{\rm out}$, an adjustable parameter varied independently of the relative magnetospheric radius $\xi$ to satisfy the inner boundary conditions: 
\begin{equation}
   \xi = \frac{R_{\rm in}}{R_{\rm A}},
   \label{eq:xi}
\end{equation}
{  where}
\begin{equation}
\displaystyle    R_{\rm A} = \left( \frac{\mu^2}{2\dot{M}_{0}\sqrt{2GM}}\right)^{2/7}
\end{equation}
is the Alfv\'{e}n radius {  (see, for instance, \citealt{elsner-lamb1977} and section 6.3 in \citealt{accrpower}).}

Because the accretion rate at the inner boundary of the disc is physically more relevant, we will also use a differently normalized version of the relative magnetospheric radius:
\begin{equation}\label{eq:xieff}
\displaystyle    \xi_{\rm eff} = R_{\rm in}\left( \frac{\mu^2}{2\dot{M}_{\rm in}\sqrt{2GM}}\right)^{-2/7} = \xi \left(\frac{\dot{M}_{\rm in}}{\dot{M}_{0}}\right)^{2/7}.
\end{equation}

\section{Results}\label{sec:res}

\subsection{Global parameters}\label{sec:xi}

The relative magnetospheric size $\xi$ is found as one of the two eigenvalues of the problem, the other being $W_{r\phi}^{\rm out}$. Other global parameters obtained in the model include several quantities at the magnetospheric boundary: the relative thickness of the disc $\left( H/R\right)_{\rm in}$, the fraction of mass reaching the magnetosphere $\dot{M}_{\rm in}/\dot{M}_{0}$, and the advection fraction $\left(Q_{\rm adv}/Q^+\right)_{\rm in }$. We also track the maximal thickness of the disc $\left( H/R\right)_{\rm max}$. 

\begin{figure}
\includegraphics[width=\columnwidth]{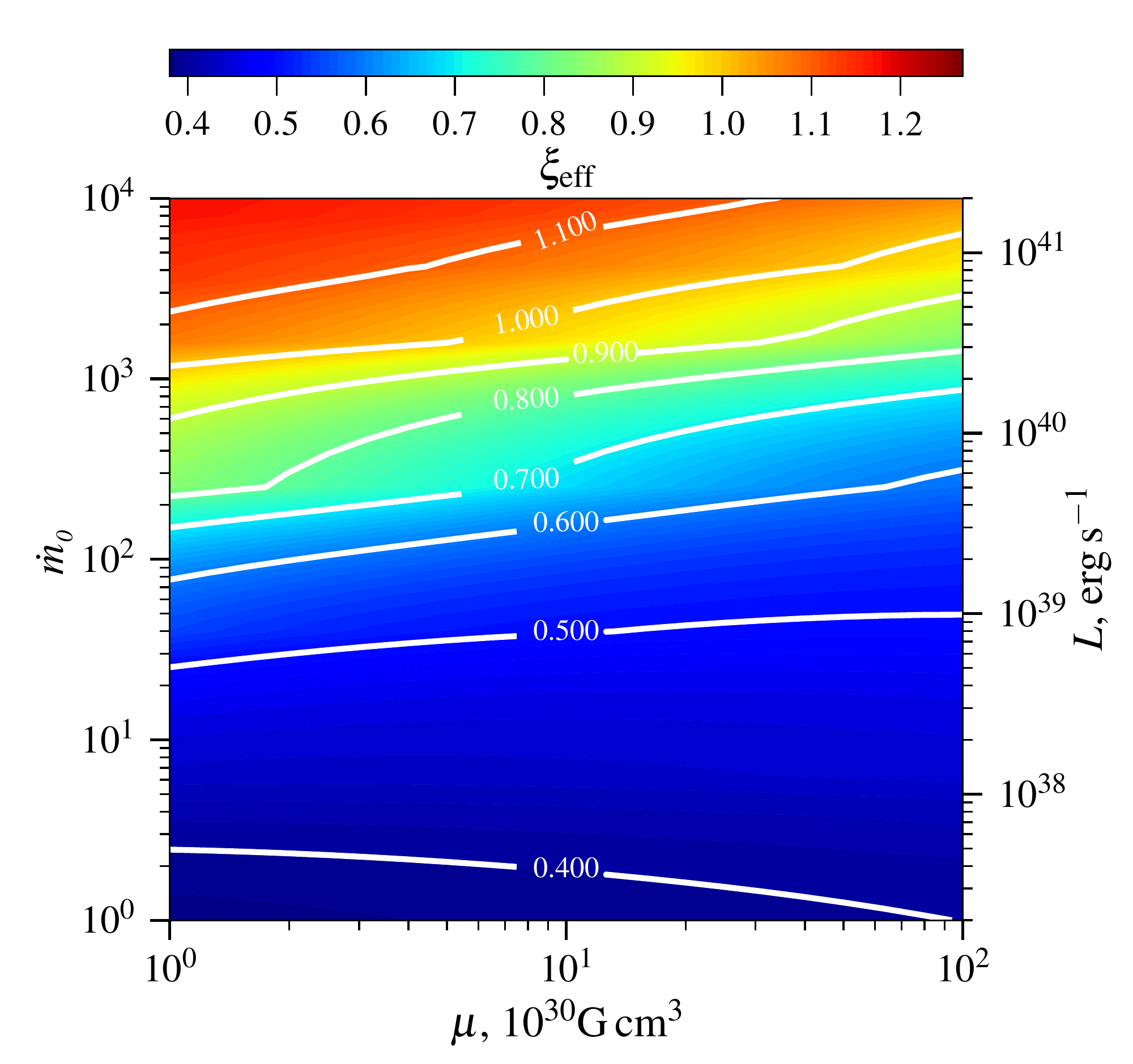}
\caption{Relative magnetospheric radius $\xi_{\rm eff}$ given by equation \eqref{eq:xieff} is shown by colour and contours on the $\mdot_0-\mu$ plane. 
The spin period is $p=10 \, p_{\rm eq}$.  
The luminosity scale is calculated assuming efficiency $\eta=0.1$, but irradiation effects on the disc structure was ignored.
}\label{fig:xi}
\end{figure}

{  All the simulations were made for the parameters: $\alpha=0.1$,  $k_{\rm t}=0.5$, $\epsilon_{\rm w}=0.5$, $\psi=1$, $n=1$ and $\eta=0$ if not stated otherwise. }
We first consider a slowly rotating accretor with the spin period equal to ten equilibrium periods $p=10p_{\rm eq}$,\footnote{When $p\gg p_{\rm eq}$, the exact value of the spin period barely affects $\xi_{\rm eff}$, see Sect.~\ref{sec:res:period}.} where the equilibrium period ({   \citealt{Lipunov92}, \citealt{Illarionov75})} is defined as the spin period for which the Alfv\'{e}n radius $R_{\rm A}$ equals to the corotation radius  $R_{\rm co}=\left(\Omega_{\rm NS}^2/GM\right)^{1/3}$ :
\begin{equation}\label{E:peq}
    p_{\rm eq} = 2\uppi (GM)^{-5/7}  \left(\frac{\mu^2}{2\sqrt{2}\dot M}\right)^{3/7}
    \simeq 1.3 m^{-5/7} \mdot_0^{-3/7} \mu_{30}^{6/7}\, {\rm s},
\end{equation}
where $\mu_{30}=\mu/10^{30}$~G\,cm$^{3}$ is dimensionless NS magnetic moment, $m=M/1.4M_\odot$ is normalized NS mass. 
We also use dimensionless mass accretion rate $\dot m_0=\dot M_0/\dot M_{\rm Edd}$ normalized by the Eddington value $\dot M_{\rm Edd}=4\uppi GM/c\kappa$. 
In Fig.~\ref{fig:xi}, we show the contours of $\xi_{\rm eff}$ in the $\mdot_0-\mu$ plane, covering two orders of magnitude in magnetic moment and five orders of magnitude in the mass accretion rate.

\begin{figure}
\includegraphics[width=\columnwidth]{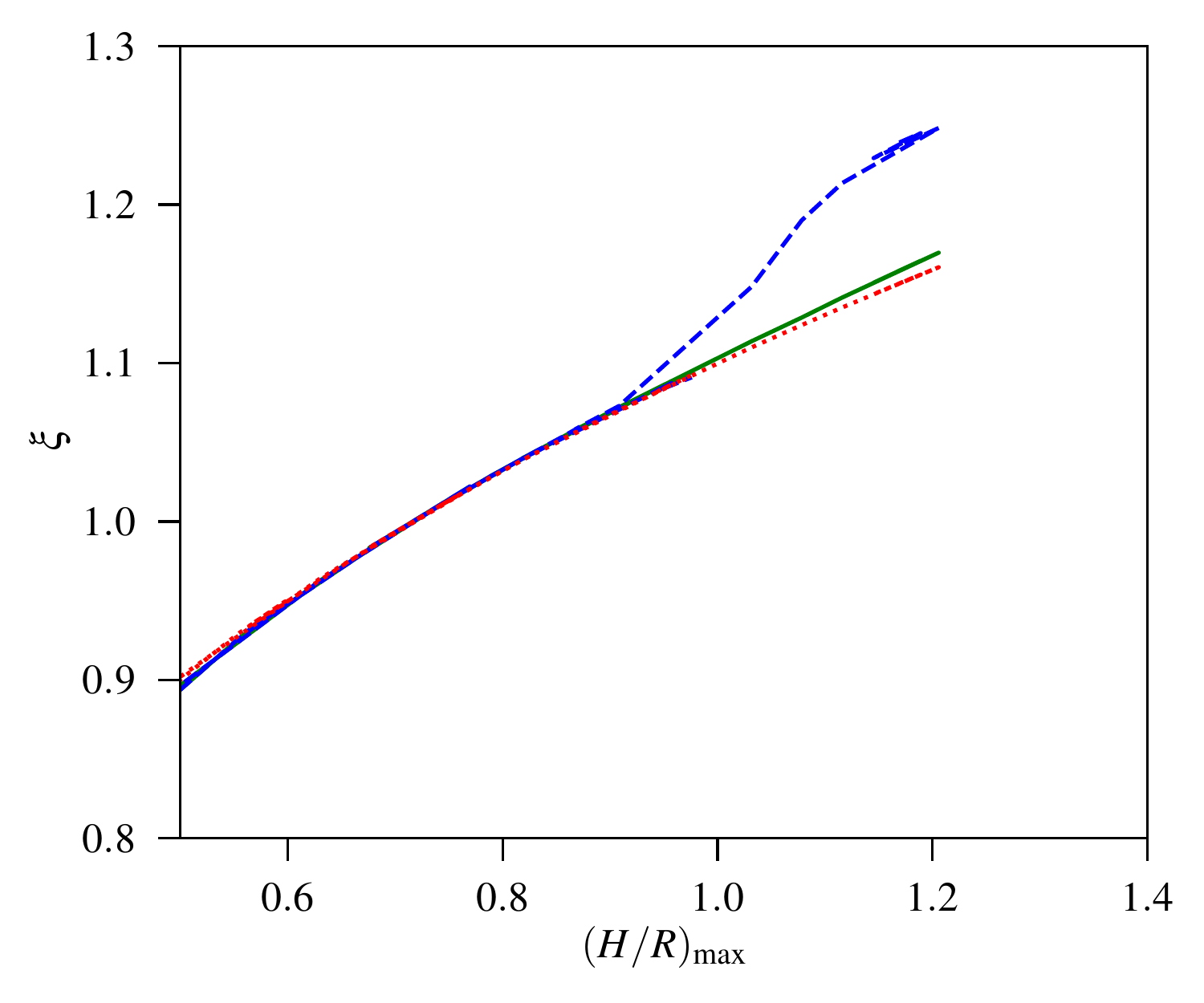}
\caption{Relative magnetospheric radii $\xi$ (dashed blue curve) and $\xi_{\rm eff}$ (green solid) as functions of the disc thickness at the magnetospheric boundary. 
The dotted red line corresponds to the long-period asymptotic given by equation (57) from \citetalias{CAP}. Magnetic moment of the NS was set to $\mu_{30}=100$. }\label{fig:xihtor}
\end{figure}

The magnetospheric radius and the thickness of the disc are tightly related. There is a good agreement with the results of \citetalias{CAP}, as we can see from  Fig.~\ref{fig:xihtor}. 
The relative magnetospheric size $\xi_{\rm eff}$ remains a monotonic function of $(H/R)_{\rm in}$ and behaves in approximate (accuracy within 5\%) accordance with equation (57) of \citetalias{CAP}. This long-period approximation works fine far from the equilibrium period, for instance, in outbursts. 

\begin{figure}
\includegraphics[width=\columnwidth]{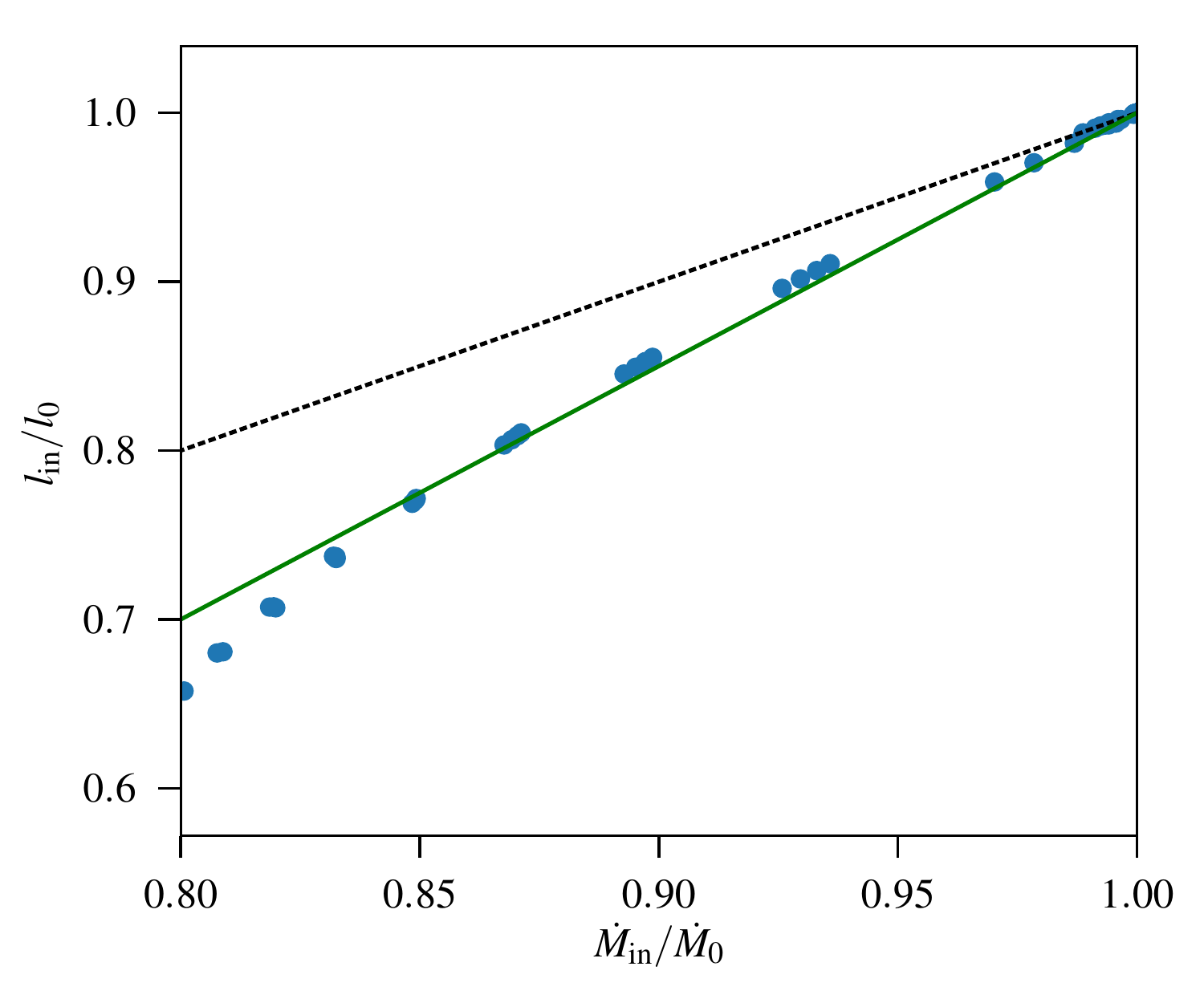}
\caption{Fraction of the angular momentum flux retained in the super-Eddington disc $l_{\rm in} / l_{0}$ as a function of the fraction of mass reaching the magnetosphere. The dotted black line corresponds to equal ratios ($l_{\rm in} / l_{0}=\dot{M}_{\rm in} / \dot{M}_{0}$, as one would expect if the net angular momentum is constant), the solid green line corresponds to the expected scaling of a growing spherization radius (equation~\ref{E:lsph}). 
}\label{fig:momentum}
\end{figure}

Not only mass but also angular momentum is lost in the wind. If there is no angular momentum flow due to stresses in the wind ($\psi=1$), the fraction of angular momentum expelled from the disc depends only on the distribution of the mass loss over the radial coordinate. The total angular momentum flux through an annulus in the disc is composed of the angular momentum carried by the matter in the disc and of the viscous torque acting on the annulus,
\begin{equation}
    l = \dot{M} \omega\sqrt{GMR}- 2\pi R^2 W_{r\phi},
\end{equation}
where $\omega=\Omega/\Omega_{\rm K}$.
Unless some of this angular momentum is removed in the wind, $l_{\rm in}=l(R_{\rm in})$ should be equal to $l_{0}=l(R_{\rm out})$. 
In Fig.~\ref{fig:momentum}, we show the ratio of the angular momentum fluxes  $l_{\rm in} / l_{0}$ as a function of the ratio of mass accretion rates $\dot{M}_{\rm in} / \dot{M}_{0}$. 
Both ratios start at unity for conservative thin disc accretion and then decrease as $\dot{M}_{0}$ increases. 
The slope of the curve in Fig.~\ref{fig:momentum} shows the evolution of the mean net angular momentum in the wind. 
With increasing mass accretion rate, the outflow involves larger radii. Characteristic radii losing most of the angular momentum is approximately equal to the spherization radius $R_{\rm sph}=\displaystyle\frac{3}{2} \dot{m}_{0} \frac{GM}{c^2}$ that leads to the scaling (see Fig.~\ref{fig:momentum}):
\begin{equation}\label{E:lsph}
    \frac{l_{0}-l_{\rm in}}{l_{\rm in } } \simeq \sqrt{\frac{R_{\rm sph}}{R_{\rm in}}} \frac{\dot{M}_{0}-\dot{M}_{\rm in}}{\dot{M}_{\rm in}}.
\end{equation}
The amount of angular momentum lost in a centrifugal wind is enhanced approximately proportionally to $\psi$.

\subsection{Dependence on the spin period}\label{sec:res:period}

The transition to the propeller regime, when the accretion flow cannot spin up the NS anymore, may be traced using the fastness parameter $\omega_{\rm s}$ defined as
\begin{equation}
\omega_{\rm s}=\displaystyle\frac{\Omega_{\rm NS}}{\Omega(R_{\rm in})}.
\end{equation}
When $\omega_{\rm s}=1$, the inner rim of the disc rotates exactly with the same frequency as the magnetosphere, making our first boundary condition marginally satisfiable.

\begin{figure}
\includegraphics[width=\columnwidth]{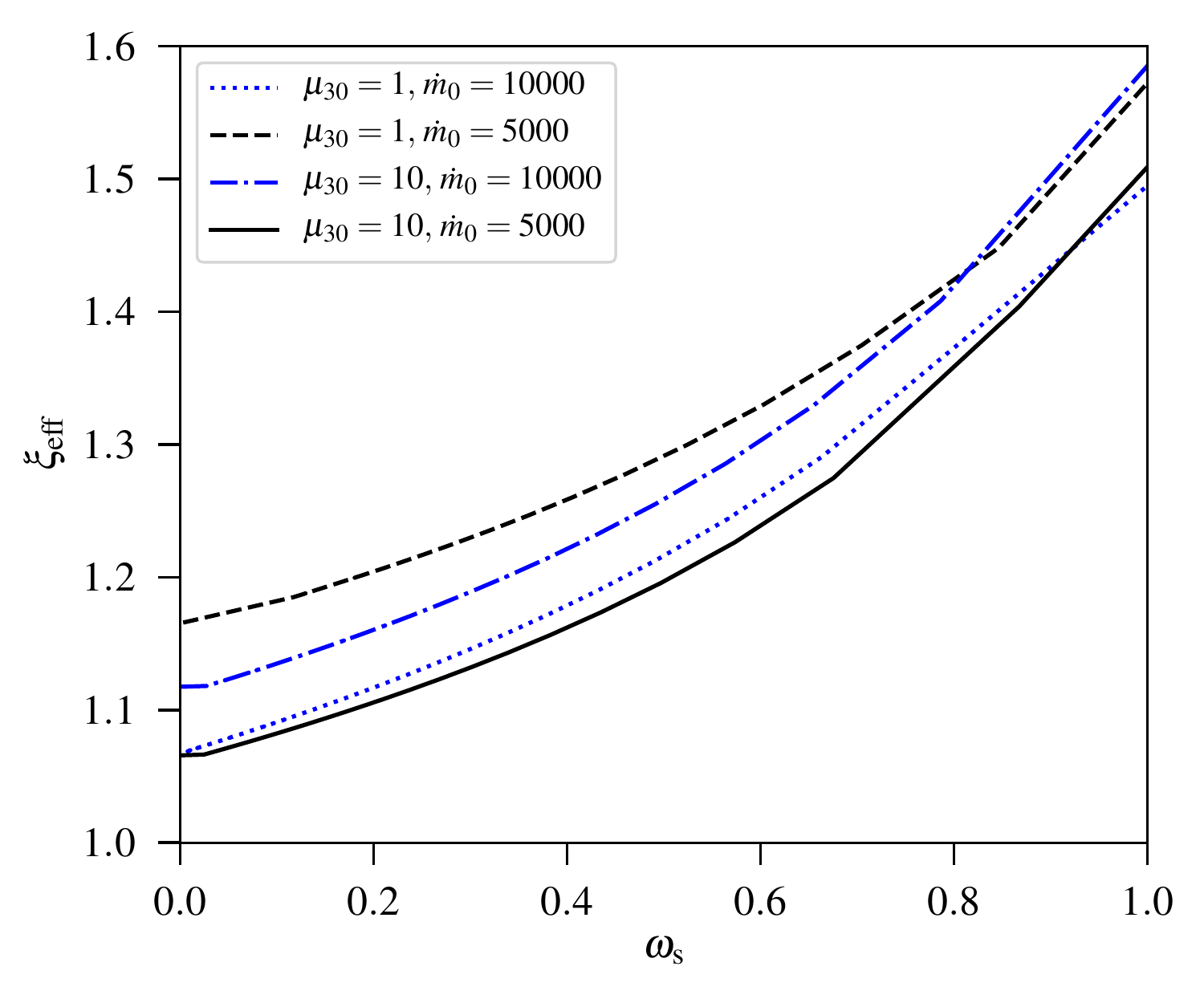}
\caption{  Parameter $\xi_{\rm eff}$ as a function of fastness  parameter for magnetic dipole moments $\mu_{30}=1$ and $\mu_{30}=10$ and mass accretion rates $\dot{m}_{0}=10^4$ and $\dot{m}_{0}=5\times 10^3$. 
}\label{fig:varps:xi}
\end{figure}

The main effect of increasing fastness parameter (or decreasing spin period of the accretor) on the properties of our solution is in the increasing {  ratio $\xi$ of the size of the magnetosphere to the Alfven radius. 
Factor  $\xi_{\rm eff}$ increases by about 40\% between the slowly rotating NS case and the propeller limit (see Fig.~\ref{fig:varps:xi}). }

Near the corotation, when $R_{\rm co}=R_{\rm in}$, disc thickness approaches zero and the boundary condition for the viscous stresses \eqref{eq:base:pbalance} reduces to the zero-torque condition used in standard disc theory. 
This allows us to compare our results directly to some of the results obtained using the codes designed for black hole accretion, e.g. by \citet{poutanen07}. 
Their model takes into account the advection effects and the outflows from the disc. 
We compare spherization radius defined as the maximal distance from the NS where the condition for outflows, $H>R$, is fulfilled.

\begin{figure}
\includegraphics[width=\columnwidth]{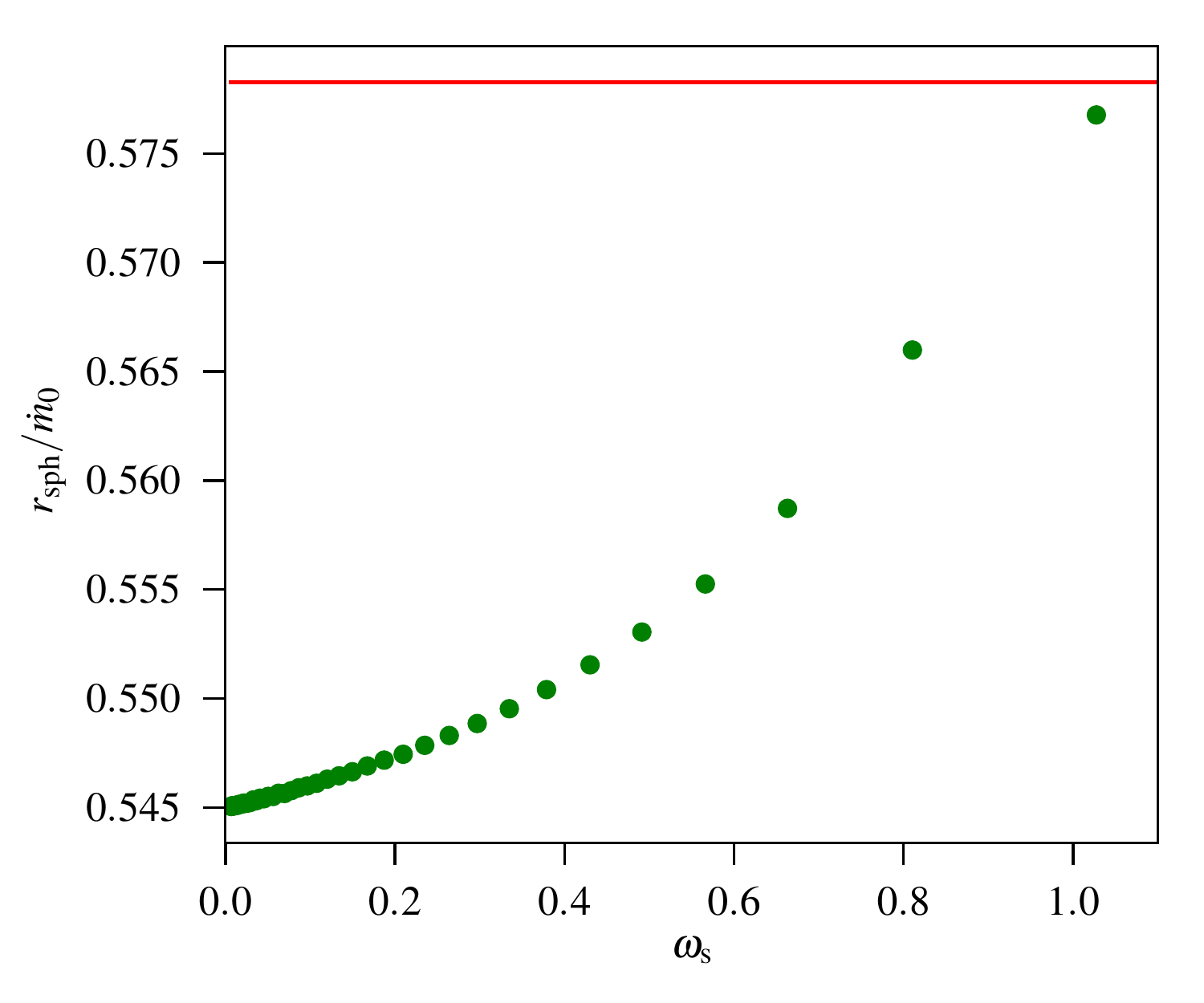}
\caption{The ratio of the spherization radius $r_{\rm sph}$ (in units of gravitational radius $R_{\rm g}$) to the dimensionless mass accretion rate $\dot{m}_{0} = 10^4$ as a function of the fastness parameter is shown by green dots. 
The red horizontal line gives the result from equation (21) of \citet{poutanen07}.
}\label{fig:varps_pout}
\end{figure}

In Fig.~\ref{fig:varps_pout}, we show the spherization radius normalized by dimensionless mass accretion rate (green dots) and compare our results to the zero-torque case. For $\epsilon_{\rm w} =0.5$ and $\mdot_{0} = 10^4$, equation (21) by \citet{poutanen07} predicts $R_{\rm sph} \simeq 0.575 \mdot_{0} ({GM}/{c^2})$, in reasonable agreement with our results close to equilibrium. 

\subsection{Effects of irradiation}\label{sec:res:irrad}

\begin{figure}
\includegraphics[width=\columnwidth]{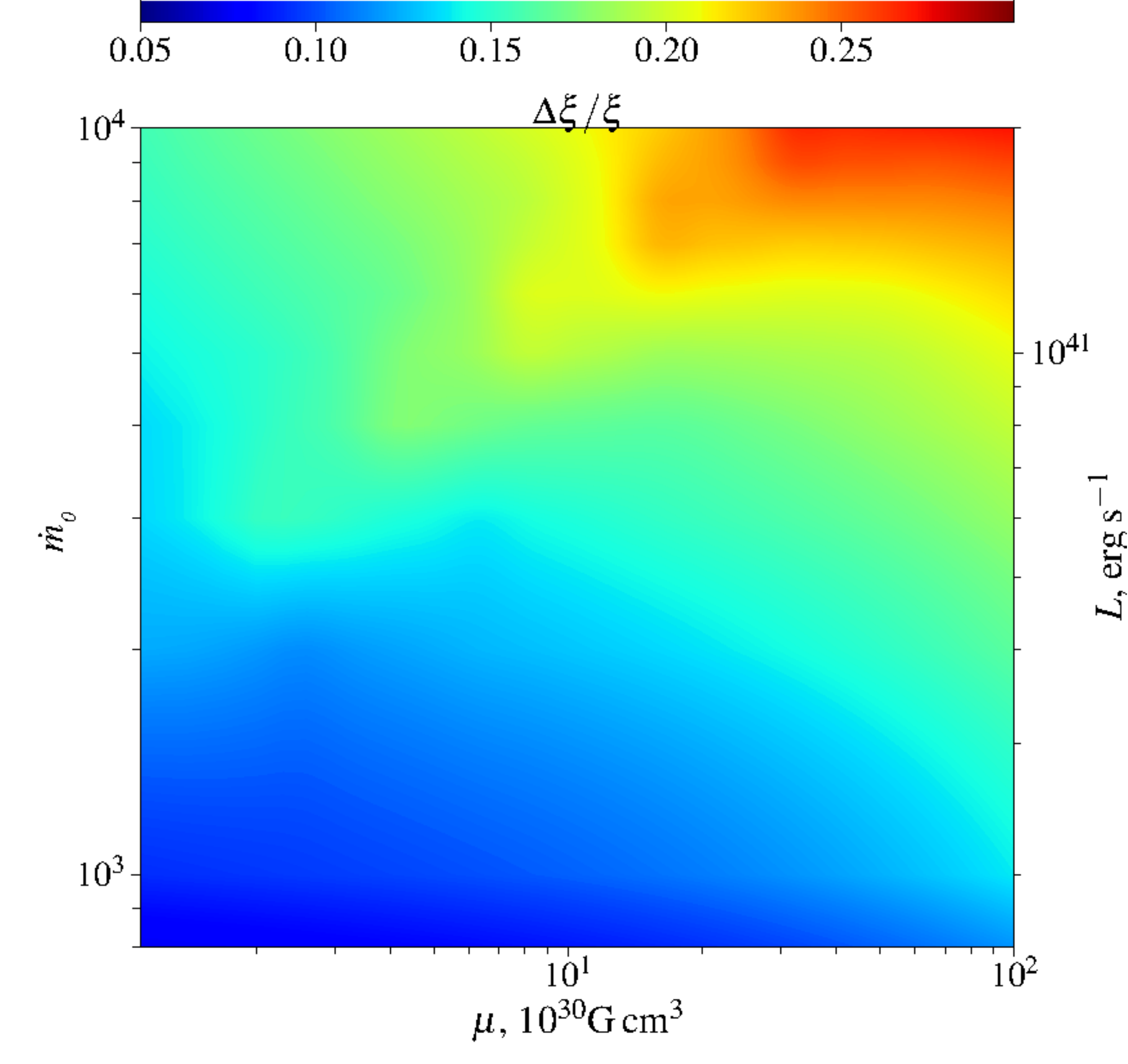}
\caption{The contours of relative correction to the magnetospheric radius caused by irradiation $\xi(\eta=0.1)/\xi(\eta=0)-1$ shown in the $\dot m_0 -\mu$ plane with colours. 
}\label{fig:eta}
\end{figure}

All the previous results were calculated without irradiation effects by setting $\eta=0$ (or $L=0$ in equations \ref{eq:base:wbalance} and \ref{eq:base:pbalance}). 
The real efficiency affecting the pressure balance is probably of the same order with the integrated accretion efficiency, though strong anisotropy of the radiation from the column is not excluded. 
Confirming the result of \citetalias{CAP}, we find that radiation from the accretion column is an important factor affecting the structure of the disc and  the radius of the magnetosphere, in particular. 
The magnetospheric radius increases by up to 30\% for $\dot m_0 \sim 10^4$, as it is shown in Fig.~\ref{fig:eta}. 
The effect grows rapidly with the mass accretion rate, as well as with the magnetic moment. The latter is a consequence of a rapidly growing ratio $P_\mathrm{rad}/P_\mathrm{mag}$ with the disc inner radius in the dipole approximation (see for instance \citetalias{CAP}, equation 63). 

\subsection{Dependence on other parameters}\label{sec:res:other}

There are adjustable parameters that influence the structure of the disc. In Table~\ref{tab:tab1} we show the global properties of the discs for different parameters. The fiducial model here is the model with $\alpha=0.1$, $\epsilon_{\rm w}=0.5$, $n=1$, $\eta=0$, and $\psi=1$. All the models are calculated for a NS with magnetic field $\mu=10^{30}$~G\,cm$^{3}$ and accretion rate $\dot m_{0}=3000$. Other models differ from the fiducial one in a single parameter.

 As in \citetalias{CAP}, we find viscosity parameter to be an important factor altering the structure of the disc. For the parameters under consideration, changing $\alpha$ from 0.1 to 0.5 results in a twofold decrease in disc thickness and $\xi$ and quenches wind formation. Changing vertical disc structure by increasing the effective polytropic index $n$ also makes the disc thinner and less likely to form outflows.

\begin{table}
  \caption{\label{tab:tab1} Disc properties.}
\begin{center}
  \begin{tabular}{lcccrr}
    \hline
     parameters & $\xi$ & $(H/R)_{\rm in}$ & $(H/R)_{\rm max}$ & $\dot M_{\rm in}/\dot M_{0}$ & $l_{\rm in}/l_{0}$ \\ \hline
     fiducial & $1.16$ & $0.97$  & $1.04$ & $0.88$ & $0.82$    \\
     $\eta=0.1$ & $1.31$ & $1.08$  & $1.09$ & $0.84$ & $0.73$ \\ 
     $\epsilon_{\rm w}=1$ & $1.18$ & $0.93$  & $1.02$ & $0.91$ & $0.88$    \\
     $\epsilon_{\rm w}=0.1$ & $1.13$ & $0.99$  & $1.06$ & $0.97$ & $0.95$    \\
     $n=1.5$ & $1.10$ & $0.92$  & $1.00$ & $0.99$ & $0.99$    \\
     $n=3$ & $1.05$ & $0.81$  & $0.91$ & $1$ & $1$    \\
     $\psi=1.5$ & $1.15$ & $0.96$  & $1.02$ & $0.91$ & $0.81$   \\
     $\alpha=0.5$ & $0.68$ & $0.61$  & $0.64$ & $1$ & $1$   \\
    \hline
  \end{tabular}
\end{center}
\tablefoot{The fiducial model has $\alpha=0.1$, $\epsilon_{\rm w}=0.5$, $n=1$, $\psi=1$, and $\eta=0$.  
  Each model differs from the fiducial one by one parameter shown in the left column. 
  All the calculations were made for $\dot m=3000$, $\mu_{30}=1$, and $p=10p_{\rm eq}$, aimed to reproduce the properties of super-critical ULXPs.}
\end{table}

\subsection{Effects of advection and wind}\label{sec:res:rastr}

\begin{figure}
\includegraphics[width=\columnwidth]{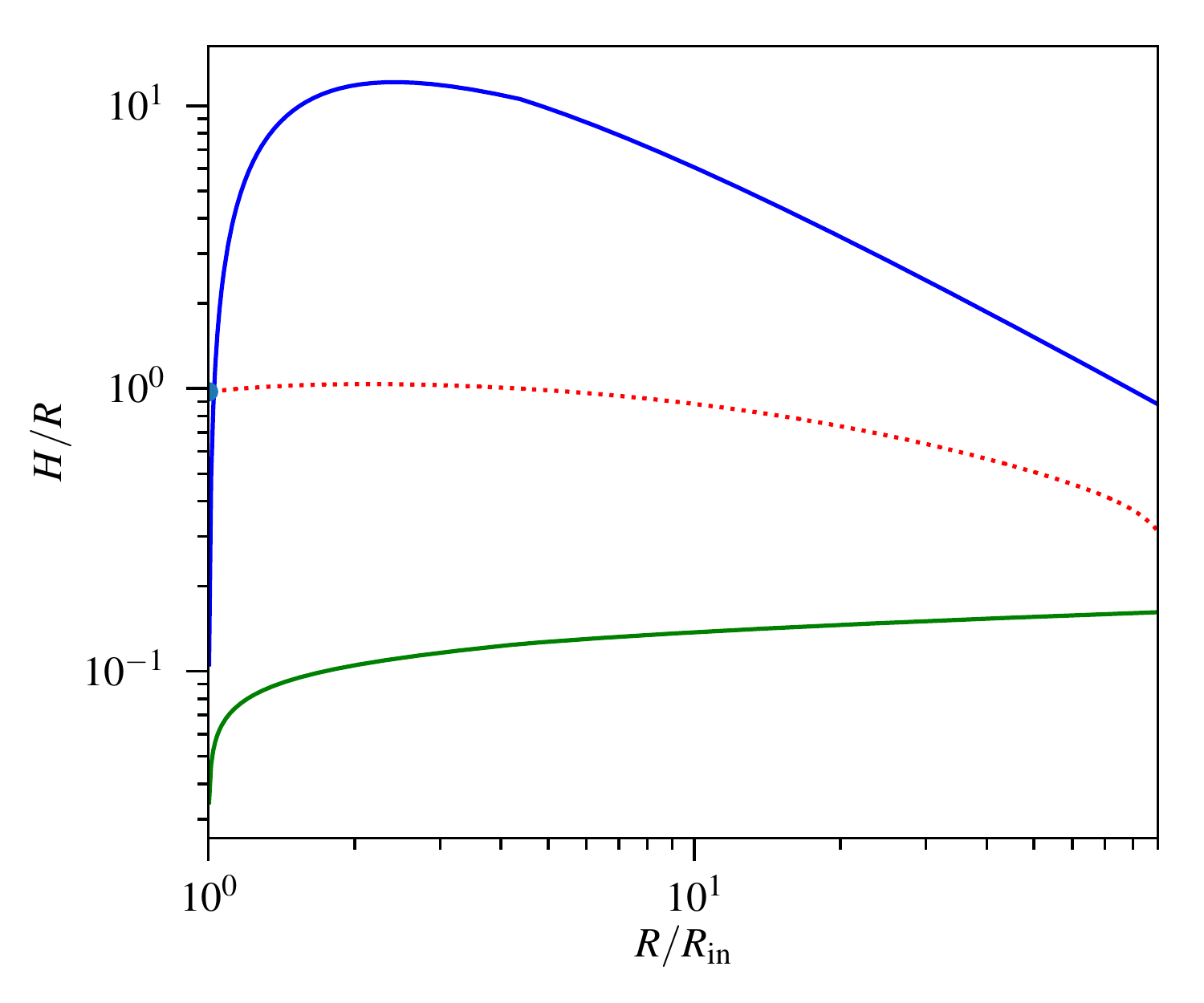}
\caption{Relative disc thickness as a function of the radius  for a model with $\mu=10^{30}{\rm G\,cm^3}$, $\mdot_0 =3\times10^3$, and $p=0.67$~s. 
Our results are shown by the red dotted curve, whereas the solid green and blue curves correspond to the asymptotics for the zones B and A  of the standard disc, respectively.  }\label{fig:hr1000}
\end{figure}

Under the assumptions we use, including the adopted vertical structure and the minimal disc thickness for wind launching, advection starts to play a role rather early, when all the disc is still sub-critical. As a consequence, the disc thickness stabilizes at $H\simeq R$ (see Fig.~\ref{fig:hr1000}). {  Slimming effect of advection was noted earlier by, for example, \citet{Slim88}, \citet{Beloborodov1998}, \citet{lipunova}, and \citet{lasota+2016}. }
The local Eddington limit ($H=R$) is reached at the critical mass accretion rate of
\begin{equation}\label{E:res:mdotcr}
\dot{m}_{\rm cr} \simeq 2000 \left( \frac{\alpha}{0.1}\right)^{2/9} \mu_{30}^{4/9},
\end{equation}
that is 5--6 times higher than in \citetalias{CAP} (equation 66). 

\begin{figure}
\includegraphics[width=\columnwidth]{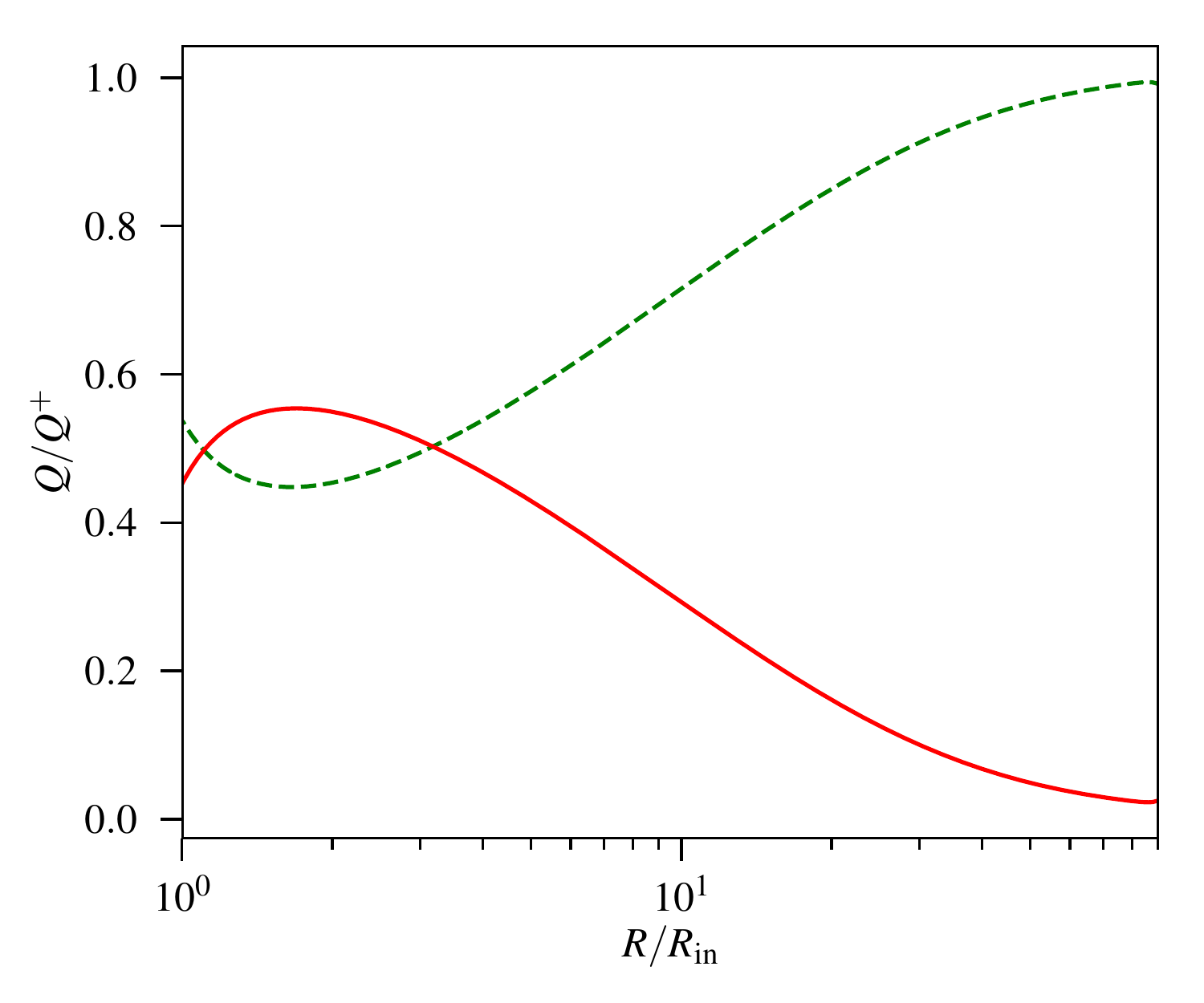}
\caption{Advection effects for the model shown in Fig.~\ref{fig:hr1000}. The advection flux fraction $Q_{\rm adv}/Q^+$ and the radiated fraction $Q_{\rm rad}/Q^+$  are shown by the red  solid and green dashed curves, respectively. 
}\label{fig:q1000}
\end{figure}

\begin{figure}
\includegraphics[width=\columnwidth]{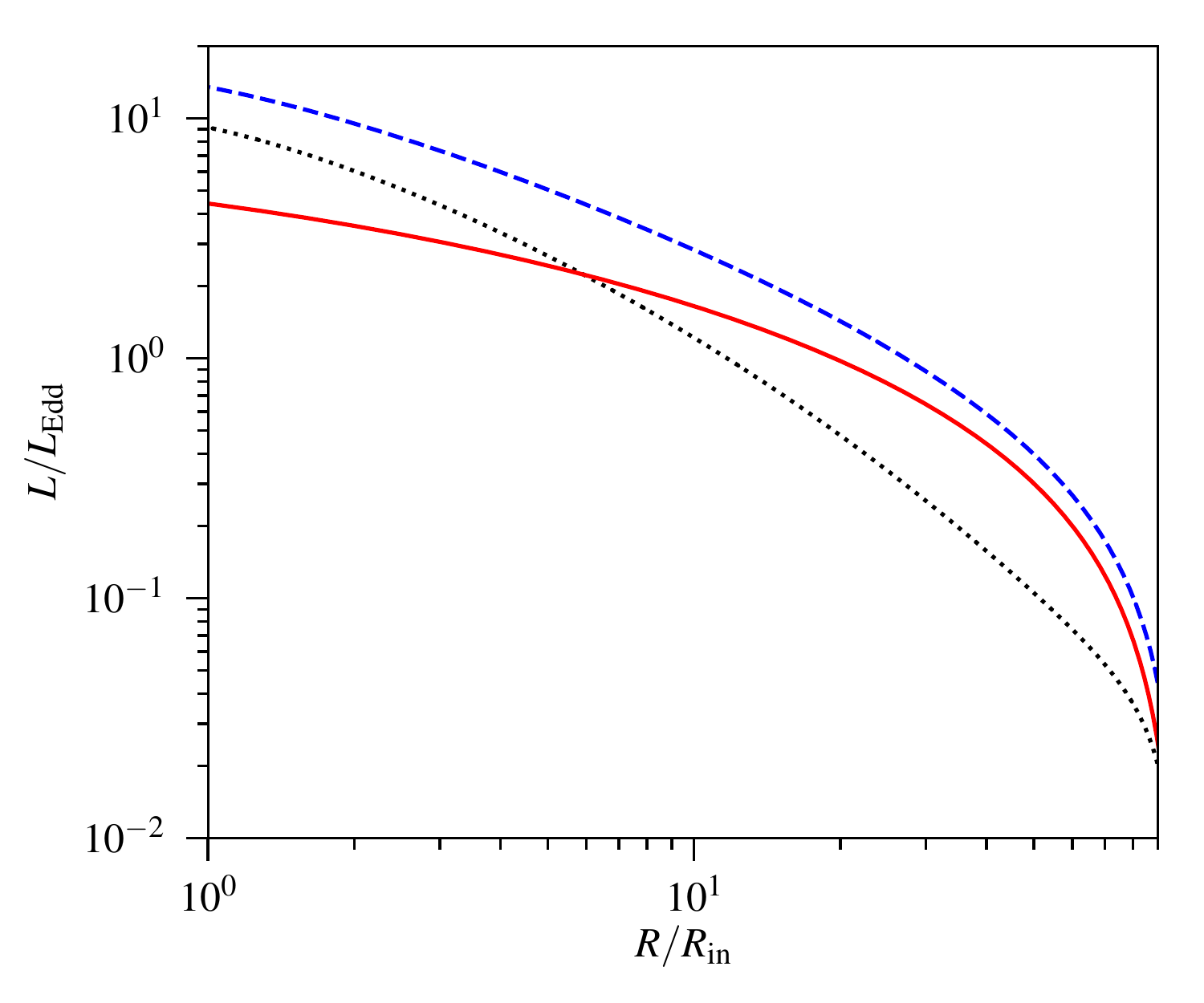}
\caption{Cumulative luminosities as functions of radius. 
The total power generated in the disc is shown by the dashed blue curve. The solid red and dotted black curves show the radiative and advected luminosities, respectively. 
Parameters of the model are $\mu_{30}=1$ and $\mdot_0 = 3000$.}\label{fig:lum3000}
\end{figure}

Advection starts to dominate in the energy balance already below this limit, at $\mdot_0 \sim 10^3$ (see Fig.~\ref{fig:q1000}), making the inner disc a huge reservoir of heat. 
For $\dot m_0$ above the limit given by equation (\ref{E:res:mdotcr}), most of the gravitational energy released in the disc is stored as heat. 
To illustrate this, we calculated the cumulative luminosity of the disc integrated from some radius $R$ to the outer radius $L_{\rm tot}=\int_{R}^{R_{\rm out}}2\pi Q^{+}RdR$.  Similarly, by integrating $Q_{\rm rad}$ and $Q_{\rm adv}$, we can define the cumulative radiative and advection powers. 
The total cumulative power is shown by a blue dashed line in Fig.~\ref{fig:lum3000}; at the inner radius $R_{\rm in}$ it is in full agreement with theoretical prediction, $L_{\rm theor}=GM\dot M/2R_{\rm in}$. 
We note that the ratio of the disc luminosity to the luminosity of the accretion column can be as low as $\sim R_{\rm NS}/R_{\rm in}$.

\begin{figure}
\includegraphics[width=\columnwidth]{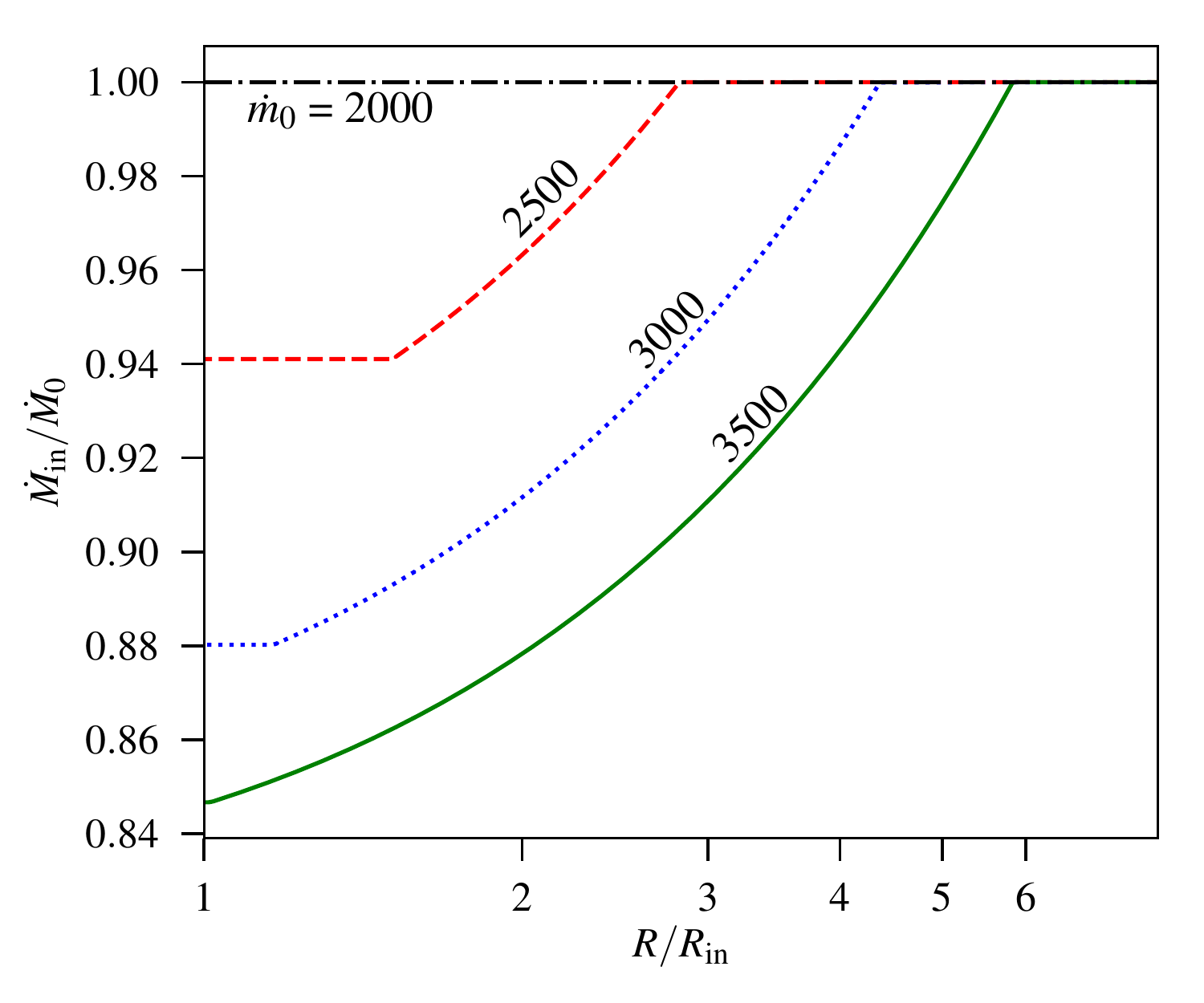}
\caption{Fraction of the mass accretion rate reaching radius $R$ for a NS with magnetic moment $\mu_{30}=1$. 
The lines from top to bottom correspond to different $\dot m_0$ in the interval 2000--3500 with step 500.  
}\label{fig:mdot}
\end{figure}

In Fig.~\ref{fig:mdot}, we plot the fraction of the initial mass accretion rate remaining in the disc for different $\mdot_0$ as a function of radius. At very high accretion rates, supercritical wind blows away a considerable part of the accretion material and operates at all radii within $R_{\rm sph}$. At some intermediate rates, $2000 \lesssim \mdot_0 \lesssim 3000$, there is a prominent sub-critical region near $R_{\rm in}$, where there is no wind. 
This is caused by the non-monotonic dependence of the disc height on radius. 
We note that amount of the blown-away material depends on the condition which is used to switch on the wind. The wind launching  relies on complex physical process and can be described only very approximately in 1-D models.

\subsection{Magnetospheric radius for different accretion regimes}

\begin{figure}
\includegraphics[width=\columnwidth]{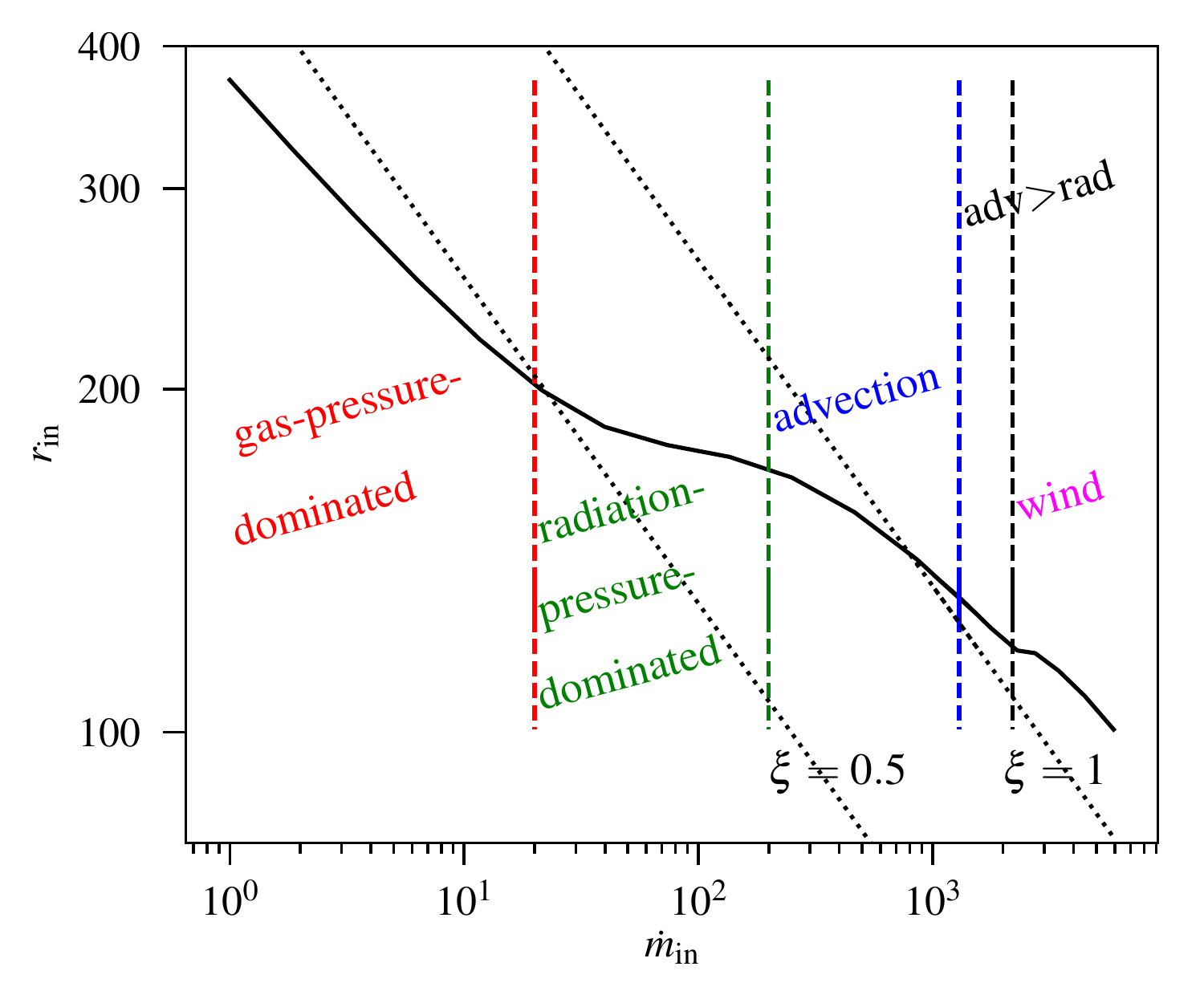}
\caption{Magnetospheric radius in units of $R_{\rm g}$  as a function of the accretion rate for a NS with magnetic moment $\mu_{30}=1$. 
Parts of the black solid curve with different slopes correspond to the different regimes of accretion near the magnetospheric boundary. 
 Two standard solutions ($r_{\rm in} \propto \mdot_0^{-2/7}$) are plotted with the grey dotted lines: $\xi=0.5$ and  $\xi=1$ (spherically-symmetric case).
}\label{fig:whole_picture}
\end{figure}

One of the most important outputs of our model is the radius of the magnetosphere. 
In many standard models \citep{GL77,K91,Wang96,KR07},  magnetospheric radius is supposed to scale with the Alfv\'{e}n radius, corresponding to constant $\xi$ in our notation. 
One of the important results of \citetalias{CAP} was understanding that the physical radius of the magnetosphere interacting with a thin radiation-pressure-dominated disc barely changes with mass accretion rate. 
Accounting for the effects of advection and wind losses makes the picture more complicated. 

Magnetospheric radius (in units of gravitational radius $R_{\rm g}=GM/c^2$) dependence on the mass accretion rate is shown in Fig.~\ref{fig:whole_picture} for a wide range of accretion rates.  The inner disc regions of most X-ray pulsars are in the gas-pressure-dominated regime. 
As the accretion rate increases, radiation pressure becomes important. 
For pulsar-scale magnetic fields, $\mu \sim 10^{30}$\,G\,cm$^3$, this happens at luminosities of a few times $L_\mathrm{Edd}$, which are quite reachable, for instance, in Be/X-ray binaries during strong outbursts like the recent super-Eddington outburst of SMC~X-3 \citep{smcx3,2017A&A...605A..39T}.  
As we have shown in \citetalias{CAP} (see eq.~61), the magnetospheric radius becomes almost independent of the accretion rate, if the radiation pressure dominates at the inner edge of a sub-critical disc:
\begin{equation}
\label{eq:rinreg}
\frac{R_{\rm in}}{R_{\rm g}}  \approx 170\, (\alpha/0.1)^{2/9}\, \mu_{30}^{4/9}\, m^{-10/9}\, .
\end{equation}
Thus, provided with a direct measurement of the magnetospheric radius, for example, from quasi-periodic oscillations, we can directly estimate the magnetic field of a NS, with a weak dependence on the viscosity parameter $\alpha$.

The inner disc radius is  defined mainly by the balance of pressures. The pressure inside the disc is related to its thickness. Thus, the dependence of $H/R$ on $\mdot_0$ is crucial for the behaviour of $r_\mathrm{in} (\mdot_0)$.  As the accretion  rate increases, advection starts to play important role. The relative thickness of the disc is no more proportional to $\mdot_0$, and the magnetospheric radius again depends on $\mdot_0$.
The interplay between wind losses and advection makes the radius dependence on mass accretion rate shallower than the $\xi = const$ approximation historically proposed for spherical accretion but stronger than $r_{\rm in} = const$.

\begin{figure}
\includegraphics[width=\columnwidth]{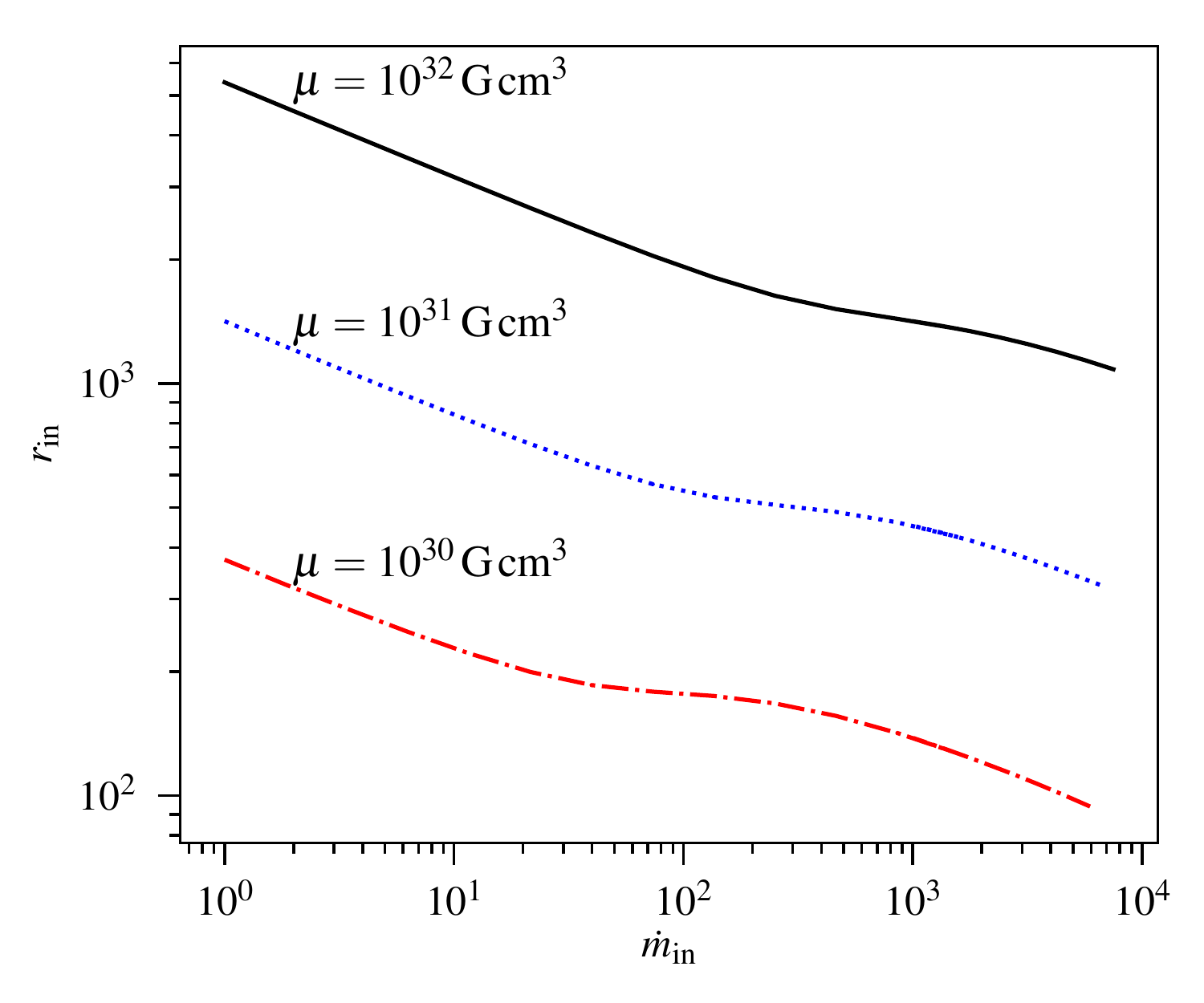}
\caption{Same as Fig.~\ref{fig:whole_picture}, but for different magnetic moments $\mu=10^{30}-10^{32}$~G\,cm$^3$.}\label{fig:whole_picture_difB}
\end{figure}

With the  increasing magnetic field, the magnetospheric radius increases, and all the boundaries between different regimes shift to higher accretion rates, as shown in Fig.~\ref{fig:whole_picture_difB}. 
In addition, the length of the plateau corresponding to the thin radiation-pressure-dominated inner disc gradually decreases with the  magnetic field, becoming effectively zero at $\mu \sim 10^{32}$G\,cm$^3$. Weakly magnetized objects, on the other hand, 
should have a prominent region of constant magnetosphere size. 
The plateau starts when radiation pressure begins to dominate over gas pressure at the radius of the magnetosphere. Position of the boundary between gas- and radiation-pressure-dominated regions of the standard disc scales with mass accretion rate as $R_{\rm ab}\propto \dot m^{16/21}$~\citep{SS73}.
Because $R_{\rm in}\propto \mu^{4/9}$  in the radiation-pressure-dominated regime (see Eq.~\ref{eq:rinreg}), the left boundary of the plateau depends on the magnetic moment as $\mdot_{\rm in, left} \propto \mu^{7/12}$.
 The right boundary of the plateau is determined by advection effects. Advection becomes important when  $Q_{\rm adv}\sim Q_{\rm rad}$, that implies $H_{\rm in}\sim R_{\rm in}$. Hence the radius at which the inner disc becomes advective scales linearly with the mass accretion rate, and the accretion rate at the right end of the plateau is $\mdot_{\rm in, right}\propto \mu^{4/9}$. 
 Thus the length of the plateau slowly decreases with magnetic moment as $\mdot_{\rm in, right}/\mdot_{\rm in, left} \propto \mu^{-5/36}$.
 
\begin{figure}
\includegraphics[width=\columnwidth]{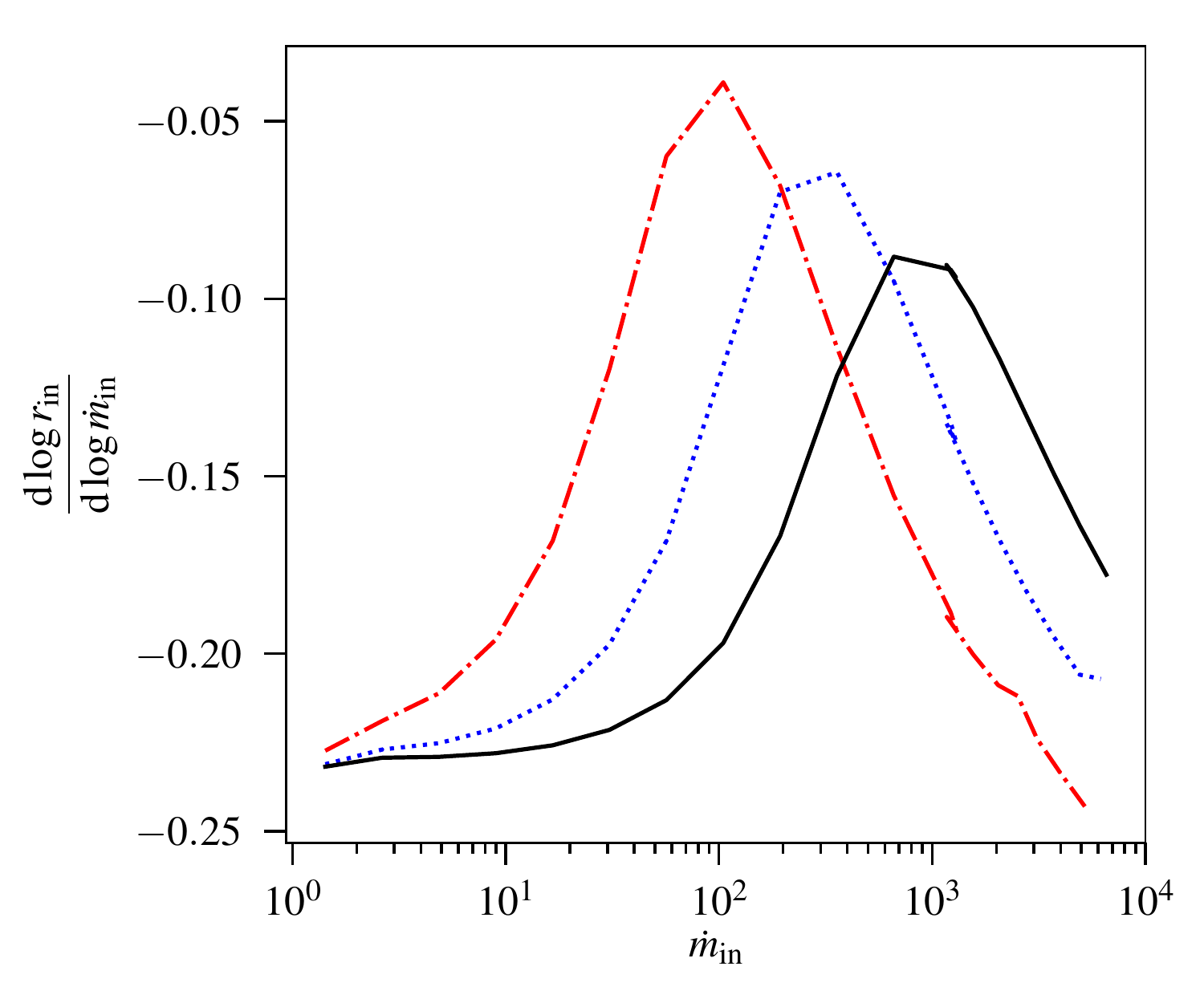}
\caption{The slope of the dependence of the magnetospheric radius on the accretion rate for models shown in Fig.~\ref{fig:whole_picture_difB}. 
}\label{fig:whole_picture_slopes}
\end{figure}

It is interesting to compare our results with the classical prescriptions \citep{GL77,Wang96}.
Fig.~\ref{fig:whole_picture} demonstrates  that the classical dependencies are much steeper, having the slope of $\delta\equiv {\rm d} \log r_{\rm in}/{\rm d}  \log {\dot m}={-2/7}$. 
In the case of $\mu=10^{30}$\,G\,cm$^{3}$, the slope is $\delta\approx-0.21$ in the gas-pressure dominated case, $\delta \approx-0.07$ in the radiation-pressure-dominated regime, and $\delta \approx-0.16$ when advection dominates. 
Evolution of the local slope $\delta$ is traced in Fig.~\ref{fig:whole_picture_slopes}. Low accretion rate asymptotic corresponding to gas-pressure-dominated thin disc stably reproduces $\delta \sim -0.23$ in accordance with the results of \citetalias{CAP} (see their Sect. 5.1). The maximal value of $\delta$ depends on how prominent is the thin radiation-pressure-dominated part of the disc for given magnetic field, changing from nearly zero for small $\mu$ to about $-0.1$ for magnetar-scale fields. Largest mass accretion rates tend to reproduce much steeper dependencies,  approaching $\delta \simeq -2/7 \simeq -0.29$.

\section{Application to ULXPs}\label{sec:discussion}

\subsection{NGC 5907~X-1 as a ULXP with a supercritical accretion disc}

The ULXP NGC\,5907~X-1\ has a huge period derivative, even after averaging in time: its period has changed from 1.43 to 1.13~s during the 10 years of observations \citep{Israel_5907}. It is also remarkable that its luminosity exceeds $10^{41}\ergl$ during some of the observations. 
The maximal detected period derivative (by absolute value) was $\dot p=-5\times10^{-9}{\rm s\,s^{-1}}$, about an order of magnitude larger than the average value for this object. The large value of $|\dot p|$ was put forward as an argument for this object being in a pure spin-up state, with the unconstrained braking term in the angular momentum equation being negligible.

The rate of change of the total angular momentum of a NS can be written as
\begin{equation}\label{eq:torques}
\displaystyle\frac{{\rm d} ( I \Omega_{\rm NS})}{{\rm d}  t}=K_{\rm su}-K_{\rm sd},
\end{equation}
where $I\simeq (1-2)\times 10^{45}{\rm \,g \,cm^{2}}$ is the moment of inertia of the NS, $K_{\rm su}=\dot M\sqrt{GMR_{\rm in}}$ and $K_{\rm sd}$ are spin-up and spin-down torques.
Ignoring the unknown spin-down contribution allows us to get a constraint
\begin{equation}\label{eq:spinup}
-\displaystyle\frac{2 \pi I \dot p}{p^2} \leq K_{\rm su}, 
\end{equation}
 resulting in an inequality for accretion rate 
\begin{equation}\label{eq:spin-up}
\dot m_{0} \gtrsim 0.105\, |\dot p|^{7/6}_{-12} p^{-7/3} I^{7/6}_{45} \mu^{-1/3}_{30}\xi^{-7/12},
\end{equation}
here $I_{45}=I/10^{45}{\rm \,g \,cm^{2}}$, and $\dot p_{-12}=\dot p/(10^{-12}$~s\,s$^{-1}$). Alternatively, having an independent estimate for the accretion rate, we then can set a lower limit for the magnetic field of this object.  {  For maximum accretion rate of $\sim 6000$, corresponding to isotropic X-ray emission, the normalised magnetic moment $\mu_{30}$ cannot be less than $\sim 0.06$.}

\begin{figure}
	\centering
	\includegraphics[width=0.5\textwidth]{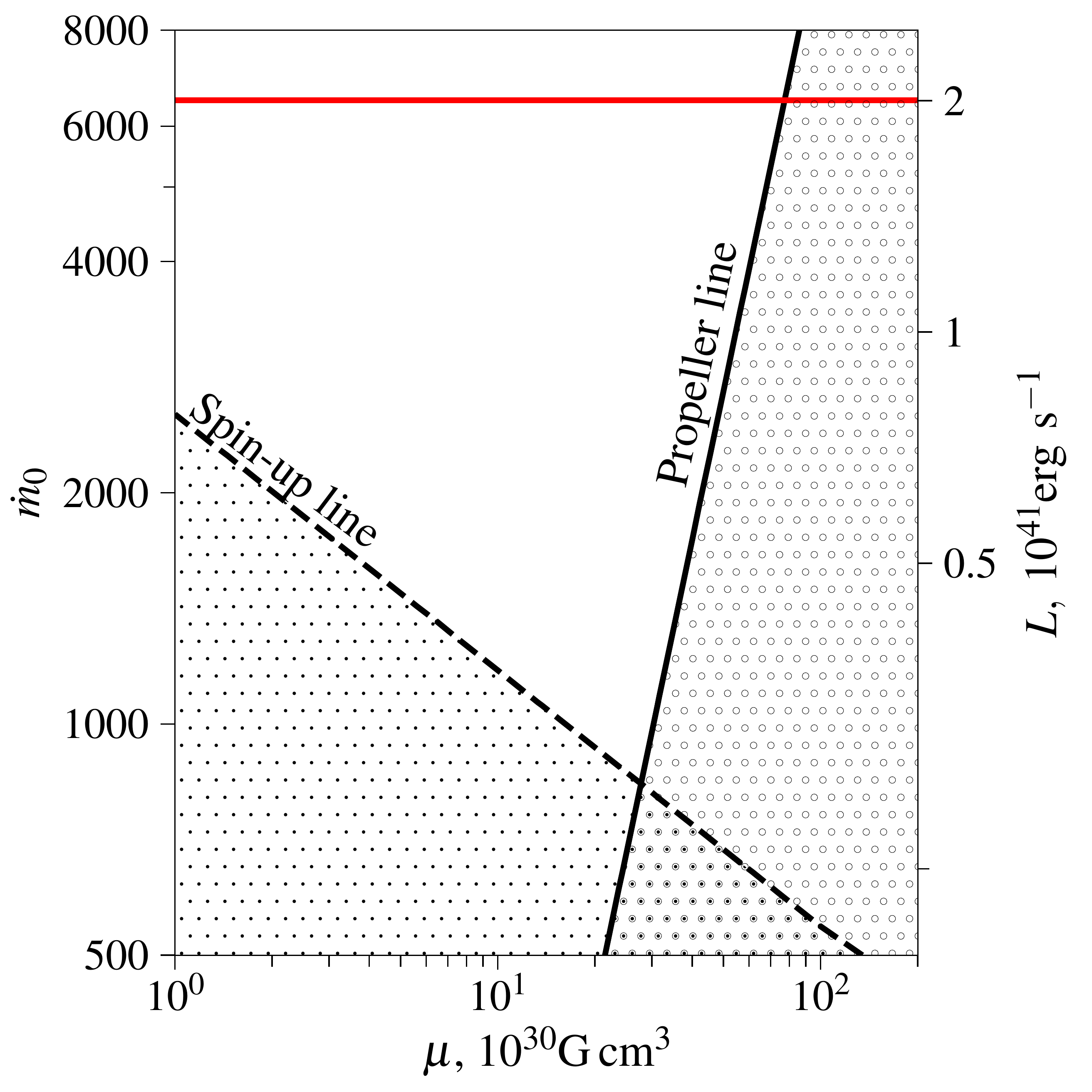}
	\caption{ 
	Restrictions for the position of NGC\,5907~X-1\ in the $\dot{m}_0 - \mu$ plane. 
	The red horizontal line corresponds to the mass accretion rate of $\dot m_0 =6500$  (accretion efficiency $\eta\simeq 0.15$). The black solid line corresponds to $p_{\rm eq} = p = 1.137$\,s. The dotted spin-up line shows the lower limit for $\dot m_{\rm in}$ set by inequality~(\ref{eq:spin-up}) using period derivation $|\dot p|=5\times 10^{-9}$~s\,s$^{-1}$. We use here $M=1.4M_{\odot}$, $R=10$~km, $I=1.5\times 10^{45}$~g cm$^2$ and $p=1.137$~s. 
	}\label{fig:dotp}
\end{figure}

In the opposite case, when a NS is close to equilibrium, its magnetospheric radius is about the size of the corotation radius.
Thus we can put an upper limit on magnetic field, suggesting the NS is still in the accretion regime. In terms of mass accretion rate, this condition may be written as  
\begin{equation}
    \dot m_{0} \gtrsim 1.8 \,\xi^{7/2} m^{-5/3} \mu_{30}^2 p^{-7/3}.
\end{equation}
Fig.~\ref{fig:dotp} shows these two limits as applied to NGC\,5907~X-1. The spin-up line  shows the lower limit for $\dot m_{0}$ set by inequality~(\ref{eq:spin-up}) using period derivative $\dot p=-5\times 10^{-9}\,{\rm s\,s^{-1}}$. 
The area below this line is forbidden unless some additional spin-up process is present. The propeller line shows the boundary of the region where the magnetospheric radius is equal to the corotational radius. The region below this line is prohibited because the accretion disc rotates slower than the magnetosphere of the NS, and no stable accretion is possible. 
Maximal bolometric luminosity of this object is about $2\times 10^{41}{\rm erg\,s^{-1}}$ \citep{Israel_5907}, that leads to a lower limit on the mass accretion rate $\mdot_{0}>  \mdot_{\rm in} \sim 6500$ assuming efficiency $\eta \sim 0.15$. 
This value of efficiency does not take into account any beaming effects that in principle can alter the observed efficiency value. 
This figure provides an evidence for a truly high mass accretion rate in NGC\,5907~X-1, as significantly low $\mdot_0 \lesssim 10^3$ are forbidden for any magnetic moments. Therefore, any beaming exceeding  a factor of several is unlikely {  because with increasing beaming, the red line in Fig.~\ref{fig:dotp} will move down, so if beaming is too large there would be no allowed region}.
 Another argument against strong beaming is the observed high pulse fractions and nearly sinusoidal profiles, inconsistent with collimation by a wind and indicating that the emission site is seen during a large fraction of the spin period.

One can set an upper limit for the magnetic field of NGC 5907 X-1 as $\mu \leq 7.5\times 10^{31}$~G\,cm$^{3}$ and $\mu \leq 5.45\times 10^{31}$~G\,cm$^{3}$ if we take into account irradiation from the column (see Sect. \ref{sec:res:irrad}).
There is an evidence for bimodal distribution in luminosities, see Fig. S2 in \citet{Israel_5907} that can be interpreted as a manifestation of the propeller effect (similarly to M82 X-2 in \citealt{Tsygankov16}). This would mean that the source is close to the propeller line. Then instead of an upper limit on the magnetic moment we get its accurate estimate. Beaming does not play a major role as magnetic field weakly depends on $\mdot_0$ (as $\mu\propto \mdot_0^{1/2}$). The disc is expected to be supercritical (i.e. having winds) in its inner parts if  $\mu_{30}\lesssim 14$ (see Eq.~\ref{E:res:mdotcr}). {  At a pulsar magnetic field $\mu_{30}\sim 1$, the expected outflow rate from the disc is about 20\% of the inflow rate, or $5\times 10^{-6}M_{\odot}\, {\rm yr}^{-1}$.}

\begin{figure}
	\centering
	\includegraphics[width=0.5\textwidth]{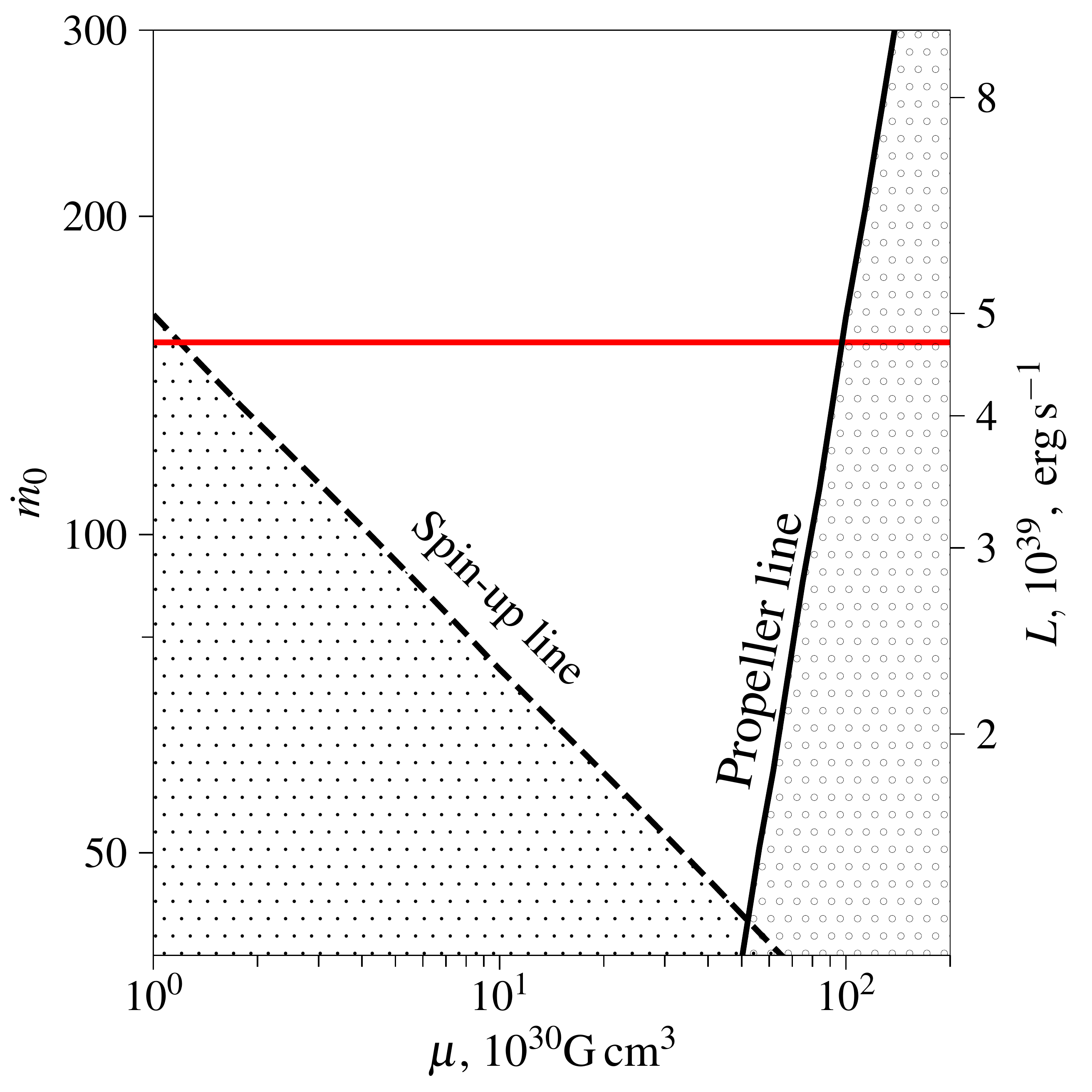}
	\caption{ 
	Restrictions for the position of NGC\,300~ULX1 in the  $\dot{m}_0 - \mu$ plane.  
	The red horizontal line corresponds to the luminosity of $4.7\times 10^{39}\ergl$ assuming accretion efficiency $\eta\simeq 0.15$.  
	 The black solid line corresponds to $p_{\rm eq} = p = 20$\,s. The dotted spin-up line  shows the lower limit for $\dot{m}_0$  set by inequality~(\ref{eq:spin-up}) using period derivation $|\dot p|=5.5\times 10^{-7}$~s\,s$^{-1}$ \citep{carpano}. We use here $M=1.4M_{\odot}$, $R=10$~km, $I=1.5\times 10^{45}$~g cm$^2$ and $p=20$~s.
	}\label{fig:pure_spinup}
\end{figure}

\subsection{NGC 300 ULX1 -- a ULXP with a strong spin-up}

The source NGC 300 ULX1 was identified as a ULXP only recently by \citet{carpano}. 
It is characterized by a moderate peak luminosity of $L\simeq 4.7\times 10^{39}\ergl$, but the observed spin-up rate of this source is exceptionally high: its spin period has changed from 45 to 20 s in less than 2 years. 
Its spin period derivative was estimated as $\dot p\simeq -5.5\times 10^{-7}$s~s$^{-1}$, that  is  the  highest $\dot p$  ever observed  from  an accreting  NS. 
Large period makes NGC 300 ULX1 a promising candidate for a NS in a pure spin-up state. 
This may be seen in Fig.~\ref{fig:pure_spinup} where the observed properties of the object are in good agreement with a pulsar-scale $\mu \sim 10^{30}$G\,cm$^3$ of a NS which is currently being rapidly spun up.
As in the case of \hbox{NGC\,5907~X-1}\ considered above, pure spin-up gives us a lower limit on the magnetic field while the propeller limit gives us an upper limit. Thus, the magnetic field of NGC 300 ULX1 is in the range $\mu=(1.5-85)\times 10^{30}$~G\,cm$^{3}$, which is not affected much by the effects of irradiation. 
The range is very wide, but sets an upper limit for the beaming factor of 2.5 only, consistent with the constraints set by \citet{Binder2018} based on the number of \ion{He}{II}-ionizing quanta.
The lower boundary here corresponds to pure spin-up and implies an equilibrium period of about $p_{\rm eq} \sim 0.2$~s. To reach such a period, the object needs to accrete at the observed rate for about $t\sim p_{\rm s}/\dot p \sim {\rm 1 yr}$. 

 At the moment of the first observations of NGC\,300\,ULX1 in 2010, the NS should have had a period even larger than $45$\,s  measured in 2016. 
The question then arises why the source had such a large period to start with.  
This period may have originated from a long episode of a very low-rate wind accretion if we assume that this NS was born rotating much faster.  
To estimate the minimal time scale of the spin-down to such a long period, we use equation (\ref{eq:torques}), neglect any spin-up torque acting on the NS and parametrize the spin-down torque  as $K_{\rm sd}= \kappa_t {\mu^2}/R^3_{\rm co}$ \citep{Lipunov92}. The time required to spin down to the observed period is about $40\,000 \, \mu_{30}^{-2}$ years.  
It means that, in order to explain a recently observed long spin period, the system with a normal, pulsar-scale magnetic field should have spent tens of thousands of years accreting at a very low rate, constantly in propeller regime. 
The very high luminosity that we observe now is an exceptionally rare event in this system. 
This is consistent with the extraordinary properties of the optical B[e] transient SN2010da this object is associated with \citep{villar16}: disappearance of a huge amount of hot dust observed before the outburst, very high peak optical luminosity, and a bright B[e] supergiant observed after the event. All this fits well with a catastrophic event in an initially broad and faint Be/X-ray binary.

\section{Conclusions}
\label{sec:conclusion}

Our model provides a simple, physically motivated description of accretion onto a NS with a magnetosphere, where the interaction between the accretion disc and the magnetosphere of the NS is reduced to a couple of boundary conditions. This allows us to reconstruct the structure of the disc 
and obtain a relative magnetospheric radius $\xi$ which is important for describing the spin evolution of magnetized NSs as well as for interpreting observational data on X-ray pulsars.  

Depending on the dipole magnetic moment of a NS and the mass accretion rate, the inner parts of the accretion disc may appear in different regimes. For classical X-ray pulsars ($\mu \sim 10^{30}{\rm \,G\,cm^3}$, $\mdot \lesssim 10$), the accretion disc remains thin and gas-pressure-dominated that implies a nearly classical scaling $R_{\rm in} \propto  \mdot^{-0.22}$. 
As the mass accretion rate increases, a large portion of the disc can exist in a radiation-pressure-dominated regime. {  Unlike in the radiation-pressure-dominated disc without advection, where the inner radius is independent of accretion rate~\citepalias{CAP}, the present model with advection yields 
$R_{\rm in} \propto  \mdot^{-(0.05-0.1)}$.}
For ULXPs, where the mass accretion rates reach $\mdot \sim 10^3-10^4$, the effects of advection  and mass loss by the wind make the scalings similar to a  spherical accretion case, resulting in a trend of about $R_{\rm in}\propto  \mdot^{-(0.2-0.3)}$. 
However, if the magnetic field is  one-two orders of magnitude larger than the usual pulsar values, 
the magnetospheric radius is larger, making the inner disc again geometrically thin and leading to a nearly flat dependence of the magnetospheric radius on $ \mdot$.

Predictions for the magnetospheric radius can be tested  through timing analysis of the stochastic component of the variability of X-ray sources, where quasi-periodic oscillations and breaks in the power-density spectra likely trace the dynamical time scales at the inner rim of the disc. Another test is spin period dynamics. We can compare the observed $\dot p$ with theoretical predictions as it was done here for NGC\,5907~X-1\ and NGC\,300~ULX1.  The constraints we get from observations of these two objects allow for a rather broad range of magnetic moments. However, the observational data set upper limits for beaming (not more than a factor of several), confirming that ULXPs are intrinsically very luminous objects rather than X-ray sources whose luminosity is amplified by an order of magnitude or more by anisotropy of their emission. 

\section*{Acknowledgements}

This research was supported by the grant 14.W03.31.0021 of the Ministry of Science and Higher Education of the Russian Federation. 
We also acknowledge support from the Russian Science Foundation grant 14-12-00146 (AC, GL, PA) and from the V\"ais\"al\"a foundation (AC). 
AC and PA thank lecturers and participants of the Astrosoma summer school\footnote{\url{http://astrosoma.ru}} who have provided an excellent environment for the development of this paper. 




\bibliographystyle{aa}
\bibliography{mybib} 



\appendix

\section{Disc vertical structure}\label{sec:vert}

For the vertical structure of the disc, we assume the form $\rho=\rho_{\rm c}(1-x^2)^n$, where $x=z/H$, and $n$ may be viewed as an effective vertical polytropic index. 
The vertical hydrostatic relation 
\begin{equation}\label{eq:vert:hydro}
\displaystyle \frac{{\rm d} P}{{\rm d} z}=-\Omega_{\rm K}^2 \,\rho\, z 
\end{equation}
implies a similar law for the vertical pressure profile $P=P_{\rm c} \left( 1-x^2\right)^{n+1}$, and a scaling relation for the disc thickness
\begin{equation}\label{eq:vert:h}
\displaystyle H = \sqrt{(2n+3)\frac{\Pi R^3}{GM \Sigma}},
\end{equation}
where the surface density and vertically-integrated pressure are related to the midplane quantities as
\begin{equation}\label{eq:vert:Sigma}
\Sigma=\displaystyle\int^{H}_{-H}\rho_{\rm c}(1-x^2)^n{\rm d}z=G_n\,\rho_{\rm c}\,H,
\end{equation}
\begin{equation}\label{eq:vert:Pi}
\Pi=\displaystyle\int^{H}_{-H}P_{\rm c}(1-x^2)^{n+1}{\rm d}z=G_{n+1}\, P_{\rm c}\, H .
\end{equation}
Here 
\begin{equation}\label{eq:vert:factorn}
G_n=\displaystyle\int^{1}_{-1}(1-x^2)^n {\rm d}x=\frac{\sqrt{\pi}\,\Gamma(n+1)}{\Gamma\left(n+\frac{3}{2}\right)}.
\end{equation}
A similar model for the vertical structure was considered by \citet{Kato08}. 
In the radiation-pressure-dominated regime $T\propto P^{1/4}$, therefore $T\propto (1-x^2)^{(n+1)/4}$. This approximation is not valid near the disc surface, because $T_{\rm eff}\neq 0$. 
The vertical radiative energy flux is determined by the vertical radiation diffusion equation:
\begin{equation}\label{eq:diffuse}
F_{\rm rad}=-D\,\nabla_z \epsilon = - D\, \dfrac{{\rm d}\epsilon}{{\rm d}z},
\end{equation}
where  $\epsilon=aT^4$ is the radiation energy density and $D=c/(3\kappa \rho)$ is the diffusion coefficient. 
We will denote the total energy release per unit surface area by $Q^+$, and the total energy leaving the two sides of the disc by  
\begin{equation}
Q_{\rm rad} = 2 \left. F_{\rm rad}\right|_{z=H} = 2\,\sigma_{\rm SB}T^4_{\rm eff}.
\end{equation}
The diffusion approximation allows us to connect the effective temperature to the temperature gradient inside the disc as
\begin{equation}
\sigma_{\rm SB}T^4_{\rm eff}=-\left. \displaystyle\frac{ca}{3\kappa \rho}\frac{{\rm d} T^4}{{\rm d} z} \right|_{z=H} . 
\end{equation}
Taking into account equations (\ref{eq:vert:Sigma}) and (\ref{eq:vert:Pi}), this expression implies
\begin{equation}\label{eq:vert:temp}
T_{\rm eff}^4=\displaystyle\frac{8}{3\kappa \Sigma}\,(n+1)\,G_n T_{\rm c}^4,
\end{equation}
and hence
\begin{equation}\label{eq:vert:qrad}
Q_{\rm rad}=\displaystyle\frac{16}{3\kappa \Sigma}\,(n+1)\,G_n \sigma_{\rm SB} T_{\rm c}^4.
\end{equation}
The midplane gas pressure is $P_{\rm g}=k\,T_{\rm c}\, \rho_{\rm c}/\tilde{m}$. The gas-to-total pressure ratio then equals to 
\begin{equation}\label{eq:vert:beta}
\beta = \frac{P_{\rm g}}{P_{\rm tot}} = \frac{kT_{\rm c} \rho_{\rm c}}{\tilde{m} P_{\rm c}} = \frac{2\,(n+1)}{2n+3} \frac{k T_{\rm c}}{\tilde{m}} \frac{\Sigma}{\Pi}, 
\end{equation}
where $\tilde{m}$ is the mean particle mass (for completely ionized gas of solar metallicity $\tilde{m}\approx 0.6m_{\rm p}$).

\section{Derivation of advection equations}\label{sec:appendix2}

The advective flux is given by expression~(\ref{eq:adv:gene}), which contains the specific (per particle) dimensionless entropy
\begin{equation}
s=\displaystyle\frac{5}{2}+\ln\left[\displaystyle\frac{3}{2}\rho^{-1}\left(\displaystyle\frac{kT}{\tilde{m}}\right)^{3/2}\right]+\displaystyle\frac{4}{3}\displaystyle\frac{\tilde{m} aT^3}{k\rho}. 
\end{equation}
The radial derivative of the entropy, under our assumptions about the vertical structure, can be written as:
\begin{equation}\label{dentropyIJK}
\displaystyle\frac{{\rm d}s}{{\rm d}R}=-\frac{{\rm d}\ln \rho_{\rm c}}{{\rm d}R}{\cal{I}}-x^2\frac{{\rm d}\ln H}{{\rm d}R}{\cal J}+\displaystyle\frac{{\rm d}\ln T_{\rm c}}{{\rm d}R}{\cal K} ,
\end{equation}
where
\begin{equation}
{\cal I}=1+\displaystyle\frac{4 \tilde{m} a T^3_{\rm c}}{3k\rho_{\rm c}}\,(1-x^2)^\frac{3-n}{4}=1+\gamma\,(1-x^2)^\frac{3-n}{4},
 \end{equation}
\begin{equation}
{\cal J}=\displaystyle\frac{5n-3}{4(1-x^2)}-\frac{\gamma}{2}\,(3-n)(1-x^2)^{-\frac{(n+1)}{4}},
 \end{equation}
\begin{equation}
{\cal K}= \frac{3}{2}+3\,\gamma\,(1-x^2)^\frac{3-n}{4},
 \end{equation}
and 
\begin{equation}
\gamma = \displaystyle\frac{4}{3} \frac{a \tilde{m} T_{\rm c}^3}{k\rho_{\rm c}}=4\,\frac{P_{\rm rad}}{P_{\rm g}} = 4\,\left(\frac{1}{\beta}-1 \right),
\end{equation}
with $\beta$ being the gas-to-total pressure ratio introduced in  equation~\eqref{eq:vert:beta}.
Vertical integration of equation~(\ref{eq:adv:gene}) allows us to write the advective flux as
\begin{equation}\label{eq:advection_term}
\displaystyle Q_{\rm adv}\!\! =\!\!-\frac{1}{2\pi (n+1)G_n}  \frac{\dot M W_{r\phi}}{R\Sigma \alpha}
\left[\frac{{\rm d}\ln \Sigma}{{\rm d}R}{\cal S}\!+\!\frac{{\rm d}\ln W_{\rm r \phi}}{{\rm d}R}{\cal P}\!+\!\frac{{\rm d}\ln T_{\rm c}}{{\rm d}R}{\cal Q}\!+\!\frac{3}{2R}{\cal R}\right], 
\end{equation}
where the dimensionless constants are
\begin{eqnarray} \label{eq:derive:S}
{\cal S}& =& \beta\displaystyle\int_{-1}^{1}(1-x^2)^\frac{5n+1}{4}\left(\frac{x^2 {\cal J}}{2}-\frac{3}{2}{\cal I}\right)\,{\rm d}x \\
& =& -\displaystyle\frac{\sqrt{\pi}(25n+9)}{16}\frac{\Gamma\left(\frac{5n+1}{4}\right)}{\Gamma\left(\frac{5n+7}{4}\right)}\beta -\frac{\sqrt{\pi}(11n+15)}{2}\frac{\Gamma\left(n+1\right)}{\Gamma\left(n+\frac{5}{2}\right)} (1-\beta) \nonumber \\
\label{eq:derive:P}
{\cal P}&=& \beta\displaystyle\int_{-1}^{1}(1-x^2)^\frac{5n+1}{4}\left(-\frac{x^2 {\cal J}}{2}+\frac{1}{2}{\cal I}\right)\,{\rm d}x \\
&=& -\displaystyle\frac{\sqrt{\pi}(5n+5)}{16}\frac{\Gamma\left(\frac{5n+1}{4}\right)}{\Gamma\left(\frac{5n+7}{4}\right)}\beta +\frac{\sqrt{\pi}(3n+7)}{2}\frac{\Gamma\left(n+1\right)}{\Gamma\left(n+\frac{5}{2}\right)} (1-\beta) \nonumber 
 , \\ 
\label{eq:derive:Q}
{\cal Q}&=&\beta\displaystyle\int_{-1}^{1}(1-x^2)^\frac{5n+1}{4}{\cal K}{\rm d}x=\frac{3\sqrt{\pi}}{2}\frac{\Gamma\left(\frac{5n+5}{4}\right)}{\Gamma\left(\frac{5n+7}{4}\right)} \beta \\ 
&+&12\sqrt{\pi} \frac{\Gamma\left(n+2\right)}{\Gamma\left(n+\frac{5}{2}\right)} (1-\beta), \nonumber  \\
\label{eq:derive:R}
{\cal R}&=&2 \,{\cal P}.
\end{eqnarray} 
To shorten the notations in equation~(\ref{eq:advection}), we also use the following combinations:   
\begin{eqnarray}\label{eq:comega}
C_{\rm \Omega}&=&\displaystyle\frac{16G_n(n+1)}{3\kappa}\frac{\sigma_{\rm SB}T^4_{\rm c}}{ \Sigma RW_{r\phi}} , \\
\label{eq:csigma}
C_{\rm \Sigma}&=&\displaystyle\frac{1}{2\pi G_{n+1}}\frac{\dot M}{\alpha R^2\Sigma}{\cal S}  , \\ 
\label{eq:cwrf}
C_{\rm wrf}&=&\displaystyle\frac{1}{2\pi G_{n+1}}\frac{\dot M}{\alpha R^2\Sigma}{\cal P}, \\ 
\label{eq:ct}
C_{\rm T}&=&\displaystyle\frac{1}{2\pi G_{n+1}}\frac{\dot M}{\alpha R^2\Sigma}{\cal Q},  \\ 
\label{eq:cfree}
C_{\rm free}&=&\displaystyle\frac{3}{4\pi R(n+1)}\frac{\dot M}{\alpha R^2\Sigma}{\cal R}  = \frac{3C_{\rm wrf}}{R}. 
\end{eqnarray}

\section{Dimensionless notation and  equations}\label{sec:notation}

Here we list the dimensionless  parameters and combinations we use throughout the paper. Our notations here are identical to those in \citetalias{CAP}. 
We normalize the NS mass as
\begin{equation}\label{all:mass}
m=\displaystyle\frac{M}{1.4{\rm M_{\odot}}}.
\end{equation}
The radius and the disc thickness   $r=R/R_{\rm g}$ and $h=H/R_{\rm g}$   are measured in units of the  gravitational radius $R_{\rm g}$. The angular frequency is normalized by the local  Keplerian frequency as
\begin{equation}\label{all:omega}
\omega=\displaystyle\frac{\Omega}{\sqrt{GM/R^3}}.
\end{equation}
The characteristic  magnetic moments of NSs lie in the range $10^{28}- 10^{32}$~G~cm$^3$, hence we normalize $\mu$ as
\begin{equation}\label{all:mu}
\mu_{30}=\displaystyle\frac{\mu}{\mu_0}=\displaystyle\frac{\mu}{10^{30}{\rm G\,cm^{3}}}.
\end{equation} 
The mass accretion rate is normalized by the Eddington value as
\begin{equation}\label{all:mdot}
\dot m=\displaystyle\frac{\dot M}{\dot M_{\rm Edd}},
\end{equation}
where
\begin{equation}\label{all:medd}
\dot M_{\rm Edd}=\displaystyle\frac{4\uppi GM}{c\kappa}\simeq2.3\times 10^{17} m \  {\rm g \, s}^{-1}.
\end{equation}
It is convenient to express the surface density in the units of the inverse opacity $\kappa^{-1}$. 
This quantity has also the physical meaning of the disc vertical optical depth 
\begin{equation}\label{all:tau}
\tau=\kappa \Sigma.
\end{equation}
The dimensionless  version of the vertically-integrated tangential stress  may be constructed as
\begin{equation}\label{all:wrf}
w_{r\phi} = \dfrac{\kappa}{c^2}W_{r\phi}.
\end{equation}
For temperatures we use the following normalization
\begin{equation}\label{all:temp}
T_{\rm c}=t_{\rm c} T_*,
\end{equation}
where
\begin{equation}\label{all:tempstar}
T_*=\left(\displaystyle\frac{GM \dot M_{\rm Edd}}{R_{\rm g}^3 \sigma_{\rm SB}}\right)^{1/4} \!\!\!\!
=\!\! \left(\displaystyle\frac{4\,\uppi \,c^5}{\kappa\, GM\, \sigma_{\rm SB}}\right)^{1/4}\!\! \!\!
\simeq 9.6\times 10^7 m^{-1/4}\,{\rm K}.
\end{equation}
The inner radius of the disc may be normalized either by the gravitational or by the Alfv\'en radius
\begin{equation}\label{all:rin}
r_{\rm in}=\displaystyle\frac{R_{\rm in}}{R_{\rm g}}=\xi r_{\rm A},
\end{equation}
where the dimensionless Alfv\'en radius is 
\begin{equation}\label{all:ra}
r_{\rm A}=\displaystyle\frac{R_{\rm A}}{R_{\rm g}}=\left(\displaystyle\frac{\lambda \mu_{30}^2}{\dot m \sqrt{2}}  \right)^{2/7} ,
\end{equation}
and
\begin{equation}\label{all:lambda}
\lambda = 
\displaystyle\frac{\mu_0^2 c^8 \kappa}{8\uppi (GM)^5} \simeq 4\times 10^{10}m^{-5}.
\end{equation}
We also introduce the natural time unit 
\begin{equation}\label{all:pstar}
p_*=\displaystyle\frac{2\uppi GM}{c^3 } \simeq 4.33\times 10^{-5}m\ {\rm s}, 
\end{equation}
that may be viewed as Keplerian rotation period at $R_{\rm g}$,
and the dimensionless  factor 
\begin{equation}\label{all:chi}
\chi=\displaystyle\frac{k}{\tilde{m}}\left(\displaystyle\frac{4\uppi}{c^3 \kappa GM \sigma_{\rm SB}}\right)^{1/4}=8.8\times 10^{-6}m^{-1/4}. 
\end{equation}
The physical meaning of $\chi$ is the square of the dimensionless speed of sound $(c_{\rm s}/c)^2$ corresponding to the characteristic temperature $T_*$. 

Here we give all the equations in dimensionless form, as they were used to calculate the disc structure.
The angular velocity at the inner boundary can be found from  equation~(\ref{eq:base:wbalance}): 
\begin{equation}\label{eq:dim:wbalance}
\omega_{\rm in}
=\displaystyle\frac{r_{\rm in}^{3/2}}{1-\eta\, h_{\rm in}/r_{\rm in}}\,\left(2\,\lambda\, \displaystyle\frac{k_{\rm t}\,\mu^2_{30} \,h_{\rm in}}{\dot m\, r^6_{\rm in}}+\displaystyle\frac{p_*}{p} \right),
\end{equation}
where $p$ is the neutron star period in seconds.
From equation~(\ref{eq:base:pbalance})  we can find the stress tensor at the boundary of the disc
\begin{equation}\label{eq:dim:pbalance}
w^{\rm in}_{r\phi}=\displaystyle
2 \,\alpha \,h_{\rm in}\,  \left(\lambda\, \displaystyle\frac{\mu^2_{30}}{r^6_{\rm in}}+\displaystyle\frac{\dot m \,\eta }{r^2_{\rm in}}\right). 
\end{equation}
The ratio of the gas pressure to the total pressure is, according to equation (\ref{eq:vert:beta}),
\begin{equation}\label{eq:dim:beta}
\beta=\displaystyle\frac{P_{\rm g}}{P_{\rm tot}}=\chi\, \displaystyle\frac{2\,(n+1)}{2n+3}\,\frac{\alpha\, \tau \,t_{\rm c}}{ w_{ r\phi}}.
\end{equation}
The differential equations (\ref{eq:dif:dwrf})--(\ref{eq:dif:sigma}) in the dimensionless form are:
\begin{eqnarray}\label{eq:dim:dwrf}
\displaystyle\frac{{\rm d} w_{\rm r\phi}}{{\rm d}r}&=&\frac{\alpha \tau }{r^2}\,(\omega^2-1), \\ 
\label{eq:dim:massacre}
\displaystyle\frac{{\rm d} \dot m}{{\rm d}r}&=&\frac{64\, \pi}{3}\,(n+1)\,G_n\frac{\epsilon_w r^2 t^4_{\rm c}}{\tau}, \\ 
\label{eq:dim:tcent}
\displaystyle\frac{{\rm d} \omega}{{\rm d}r}&=&-\frac{1}{2}\frac{\omega}{r}+\frac{\alpha \tau}{2\dot m r^{1/2}}\,(\omega^2-1) \nonumber \\  &+& \frac{64\,\pi}{3}G_n\,(n+1)\,\epsilon_w\, (\psi-1) \frac{\omega\, t^4_{\rm c}r^2}{\tau \dot m}+\frac{w_{\rm r \phi} r^{1/2}}{\dot m}, \\ 
\label{eq:dim:omega}
\displaystyle\frac{{\rm d} t_{\rm c}}{{\rm d}r}&=&\frac{t_{\rm c}}{8-6\,\beta}\,\left(\frac{1-3\,\beta}{\tau}\,\displaystyle\frac{{\rm d} \tau}{{\rm d}r} \right. \nonumber \\ 
&+& \left. \frac{\alpha \tau}{w_{\rm r\phi}r^2}(\omega^2-1)(1+\beta)-\frac{3(1-\beta)}{r} \right) . 
\end{eqnarray}
Taking into account the sign of $d\Omega / dR < 0$, we re-write equation (\ref{eq:dif:sigma}) as
\begin{eqnarray}
\label{eq:dim:sigma}
\displaystyle\frac{{\rm d} \ln \tau}{{\rm d}r}&=&\left[C^*_{\rm \Sigma}+\frac{C^*_{\rm T}}{8}\frac{1-3\beta}{1-\frac{3}{4}\beta} \right]^{-1}
\left[C^*_{\rm \Omega}-\frac{3}{2}\frac{\omega}{r^{5/2}}+\frac{1}{r^{3/2}}\displaystyle\frac{{\rm d} \omega}{{\rm d}r}  \right. \nonumber \\ 
&-&  \left. \frac{\alpha \tau}{w_{\rm r\phi}r^2}(\omega^2-1)\left(C^*_{\rm wrf}+\frac{C^*_{\rm T}}{8}\frac{1-3\beta}{1-\frac{3}{4}\beta} \right) \right. \nonumber  \\
&+&  \left. \displaystyle\frac{3 C^*_{\rm T}}{8r}\frac{1-\beta}{1-\frac{3}{4}\beta}-C^*_{\rm free} 
\right], 
\end{eqnarray}
where the dimensionless versions of coefficients (\ref{eq:comega})--(\ref{eq:ct}) are
\begin{eqnarray}
C^*_{\rm \Omega}&=&\displaystyle\frac{64\,\pi}{3}\,G_n\,(n+1)\,\frac{t^4_{\rm c}}{\tau\, r\, w_{r\phi}}\, , \\
C^*_{\rm \Sigma}&=&\frac{2}{G_{n+1}}\,\frac{\dot m}{\alpha\, r^2\,\tau}\,{\cal S}\, , \\
C^*_{\rm wrf}&=&\frac{2}{G_{n+1}}\frac{\dot m}{\alpha\, r^2\,\tau}\,{\cal P}\, , \\
C^*_{\rm T}&=&\frac{2}{G_{n+1}}\,\frac{\dot m}{\alpha\, r^2\,\tau}\,{\cal Q}\, .
\end{eqnarray}


\label{lastpage}
\end{document}